\documentclass[aps,prb,twocolumn,superscriptaddress]{revtex4-2}
\usepackage{graphicx,color}
\usepackage{amsthm}
\usepackage{amsfonts}
\usepackage{algorithmic}
\usepackage{enumerate}
\usepackage{latexsym}
\usepackage{amsmath}
\usepackage{amssymb}
\usepackage{subfig}
\usepackage{adjustbox}
\usepackage[section]{placeins}
\usepackage{float}
\usepackage{ragged2e}
\usepackage{etoolbox} 

\usepackage[colorlinks=true,citecolor=blue,linkcolor=blue]{hyperref}

\usepackage{dcolumn}
\usepackage{bm}


\usepackage{caption}
\usepackage[T1]{fontenc}
\usepackage{mathptmx}

\makeatletter
\def\fps@figure{htbp}
\def\fps@table{htbp}
\makeatother

\begin{document} 
\title{Tuning superconducting pairing symmetry via a staggered potential in the doped honeycomb Hubbard model}
\author{Yanmei Cai}
\affiliation{School of Physics and Astronomy, Beijing Normal University, and Key Laboratory of Multiscale Spin Physics (Beijing Normal University), Ministry of Education, Beijing 100875, China\\}
\author{Yicheng Xiong}
\affiliation{School of Physics and Astronomy, Beijing Normal University, and Key Laboratory of Multiscale Spin Physics (Beijing Normal University), Ministry of Education, Beijing 100875, China\\}
\author{Ying Liang}
\affiliation{College of Physics, Hebei Normal University, and Hebei Advanced Thin Films Laboratory, Shijiazhuang 050024, China\\}
\affiliation{School of Physics and Astronomy, Beijing Normal University, and Key Laboratory of Multiscale Spin Physics (Beijing Normal University), Ministry of Education, Beijing 100875, China\\}
\author{Tianxing Ma}
\email{txma@bnu.edu.cn}
\affiliation{School of Physics and Astronomy, Beijing Normal University, and Key Laboratory of Multiscale Spin Physics (Beijing Normal University), Ministry of Education, Beijing 100875, China\\}

\begin{abstract}
The ability to control superconducting pairing symmetry is crucial for designing unconventional and topological superconductors, yet practical tuning parameters beyond chemical doping remain limited. In this study, we investigate the effect of a tunable sublattice staggered potential on the pairing symmetry in the doped honeycomb Hubbard model. Determinant quantum Monte Carlo at finite temperature and constrained-path quantum Monte Carlo at zero temperature are employed to compute spin susceptibilities and pairing correlations in different channels. We find that increasing the staggered potential suppresses antiferromagnetic fluctuations and, at low doping, induces a transition in the dominant pairing tendency from $d+id$-wave to $f_n$-wave, with consistent results from both quantum Monte Carlo methods. In contrast, at higher doping levels, the system remains dominated by $d+id$-wave pairing even under an enhanced staggered potential. Moreover, strengthening the on-site interaction $U$ enhances the dominant pairing channel, underscoring the essential role of electronic correlations. Our results establish the staggered potential as a practical band-engineering tool for selecting unconventional pairing symmetries without varying the doping concentration, providing inspiration for designing graphene-based artificial superconductors and related doped band insulators such as Li${}_x$MNCl.
\end{abstract}

\maketitle
\section{Introduction}
\label{sec:I} 

Spin-triplet unconventional superconductivity is of great interest for quantum information due to its potential to host topological excitations such as Majorana zero modes \cite{PhysRevB.61.10267,zheng_high_2023}. Nevertheless, candidate materials exhibiting clear signatures of such states remain scarce, and their underlying pairing mechanisms are still actively debated \cite{RevModPhys.75.657,doi:10.1126/science.aav8645,contamin_hybrid_2021}. Recently, a promising and tunable platform has emerged in lightly doped band insulators, such as electron-doped \(\mathrm{SrTiO}_3\), \(\mathrm{Li_{x}MNCl}\), which exhibit unconventional superconductivity alongside intriguing properties \cite{GASTIASORO2020168107,PhysRevLett.97.107001,PhysRevLett.103.077004,PhysRevB.96.024518}. A central and unresolved challenge in this field is the controlled design and manipulation of superconducting order parameters, particularly their pairing symmetry. While chemical doping has been the primary experimental knob for tuning superconductivity, it often introduces disorder \cite{tanaka_superconductivity_2022}. Therefore, the pursuit of more flexible tuning knobs is of great importance.

Recently, Crépel and Fu proposed an intriguing theoretical framework providing a unified explanation for unconventional superconductivity in doped band insulators \cite{doi:10.1073/pnas.2117735119}. In a honeycomb Hubbard model with a sublattice potential difference, Coulomb repulsion can induce effective attraction through virtual interband transitions or excitonic processes, thereby stabilizing spin-triplet pairing. This framework predicts phenomena such as a direct superconductor-insulator transition  and a Bose-Einstein condensate to Bardeen-Cooper-Schrieffer crossover at low densities \cite{doi:10.1073/pnas.2117735119,crepel_new_2021}. However, these results primarily rely on analytically controllable expansions and mean-field treatments. In the intermediate and strong coupling regime of the Hubbard model, where strong correlations dominate, quantum fluctuations and competition among multiple pairing channels may influence the selection of pairing symmetries. Therefore, unbiased and rigorous numerical verification is urgently required. Motivated by this, we employ two kinds of quantum Monte Carlo (QMC) approaches that can capture the quantum fluctuations beyond mean-field theory to conduct a systematic investigation of spin fluctuations and various pairing correlations in the strongly correlated doped honeycomb Hubbard model with a tunable sublattice staggered potential $\Delta$. As nonperturbative numerical techniques, they are well suited for computing magnetic correlations in Hubbard models with on-site interactions \cite{PhysRevLett.110.107002,PhysRevB.94.075106}.

Although intrinsic graphene exhibits semimetallic behavior near the Dirac point, the introduction of a nonzero sublattice staggered potential $\Delta$ between the A and B sublattices opens a band gap, transforming the system into a band insulator with a tunable gap. This establishes the staggered potential not merely as a model parameter, but as a key physical degree of freedom that can fundamentally alter the electronic structure. Experimentally, such a symmetry-breaking field can be intrinsically present in certain materials or controllably induced via external gate voltages or through interfacial engineering in van der Waals heterostructures, making it a highly flexible tool for material design \cite{wu_electrostatic_2023}. 

In this context, electric field control of superconductivity provides a clear and direct illustration: experiments on twisted double bilayer graphene \cite{shen_correlated_2020}, bilayer graphene \cite{PhysRevLett.99.216802,zhang_direct_2009} and magic-angle twisted trilayer graphene systems \cite{2021Sci...371.1133H,lui_observation_2011,zhou_half-_2021} have demonstrated that superconductivity can be tuned by an applied displacement field. Corresponding theoretical studies have employed QMC numerical methods to systematically investigate correlation effects and pairing tendencies in trilayer graphene under applied electric fields, providing theoretical support for electric-field-controlled correlated superconductivity \cite{PhysRevB.107.245106}. A similar electric field tuning approach has recently been extended to nickelate platforms, where Yang \(et\) \(al\). proposed that a realistic vertical electric field could enhance superconductivity in monolayer and bilayer \(\mathrm{La_{3}Ni_{2}O_{7}}\) thin films \cite{shao_possible_2026}. In systems with a buckled honeycomb lattice, such as silicene, a vertical electric field can generate an effective staggered potential term due to the height difference between the two sublattices, offering a clear material realization of the correspondence between electric field and staggered potential \cite{PhysRevLett.109.055502,jx2x-fb5b}.

Previous studies have shown that \(d\)-wave superconductivity may emerge in lightly doped honeycomb Hubbard models \cite{nandkishore_chiral_2012,PhysRevB.78.205431,PhysRevB.86.020507}, 
while other works suggest a competition between various chiral channels, such as chiral \(d+id\)-wave and  \(p + ip\)-wave states, with the outcome sensitive to specific model parameters and doping levels
\cite{PhysRevB.90.054521,PhysRevB.94.115105,PhysRevLett.98.146801,PhysRevB.92.085121,PhysRevB.102.125125,PhysRevLett.100.146404,PhysRevB.105.L100505}. {Despite this progress, a systematic investigation of the role of the staggered potential at fixed doping, particularly how it modulates magnetic fluctuations and superconducting pairing symmetries, remains lacking.} \color{black} A clear understanding of this interplay is essential for achieving controlled symmetry engineering in synthetic quantum materials, and could bridge theoretical proposals with feasible experimental tuning knobs in platforms such as graphene heterostructures, twisted bilayer systems, and engineered atomic lattices.

In this work, by means of two complementary large-scale QMC methods, we systematically investigate the Hubbard model on the honeycomb lattice in the presence of a tunable staggered sublattice potential $\Delta$: finite-temperature determinant quantum Monte Carlo (DQMC) method \cite{PhysRevD.24.2278,PhysRevB.31.4403} and zero-temperature constrained-path quantum Monte Carlo (CPMC) method \cite{PhysRevB.55.7464,PhysRevB.63.115112}. A key advantage of using $\Delta$ as a control parameter is that it offers a conceptually clean pathway to alter electronic correlations without changing the chemical composition or introducing extrinsic disorder. 
We compute the spin susceptibility and pairing correlation functions for various channels, including the \(d+id\)-wave and high-angular-momentum \( f_n \)-wave channels. Our key finding is that at small doping levels, increasing the staggered potential suppresses antiferromagnetic fluctuations and drives an evolution of the dominant superconducting pairing symmetry from \(d+id\)-wave to \( f_n \)-wave, with both QMC methods yielding consistent results. 

The paper is organized as follows: Sec. \ref{sec:II} introduces the model and numerical methods, Sec. \ref{sec:III} presents the results on pairing symmetry evolution, and Sec. \ref{sec:IV} summarizes the conclusions.

\section{Model and Methods}
\label{sec:II}
\vspace{0.5cm}
We consider the Hubbard model on the honeycomb lattice with staggered potential on A/B sites and on-site interactions. The Hamiltonian takes the form \cite{PhysRevB.70.195122,PhysRevB.72.085123,PhysRevB.73.125411,RevModPhys.81.109} 
\begin{align}
H = & -t \sum_{i\eta\sigma} \left( a_{i\sigma}^\dagger b_{i+\eta,\sigma} + \text{h.c.} \right) + \frac{\Delta_0}{2} \left[ \sum_{i\in B} n_{b,i} - \sum_{i\in A} n_{a,i} \right] \nonumber\\
& + U_A \sum_{i\in A} n_{a,i\uparrow} n_{a,i\downarrow} + U_B \sum_{i\in B} n_{b,i\uparrow} n_{b,i\downarrow} + \mu \sum_{i\sigma} \left( n_{a,i\sigma} + n_{b,i\sigma} \right).
\label{eq:model1}
\end{align}
Here, \( a_{i\sigma} \) (\( a_{i\sigma}^\dagger \)) annihilates (creates) an electron at site \( \mathbf{R}_i \) with spin \( \sigma \) (\( \sigma = \uparrow, \downarrow \)) on sublattice A, and \( b_{i\sigma} \) (\( b_{i\sigma}^\dagger \)) annihilates (creates) an electron at site \( \mathbf{R}_i \) with spin \( \sigma \) (\( \sigma = \uparrow, \downarrow \)) on sublattice B. The number operator is defined as \( n_{a,i\sigma} = a_{i\sigma}^\dagger a_{i\sigma} \). \( t \) is the nearest-neighbor hopping integral, \( \Delta_0 \) is the staggered sublattice potential, \( U_A \) and \( U_B \) are the on-site Hubbard interactions, and \( \mu \) is the chemical potential. In our simulations, we mainly set \( U_A \) = \( U_B \) = \( U = 3t \), corresponding to an intermediate electron correlation strength.

\begin{figure}[!htbp]
\vspace{0.2cm}
\centering
\includegraphics[width=0.48\textwidth]{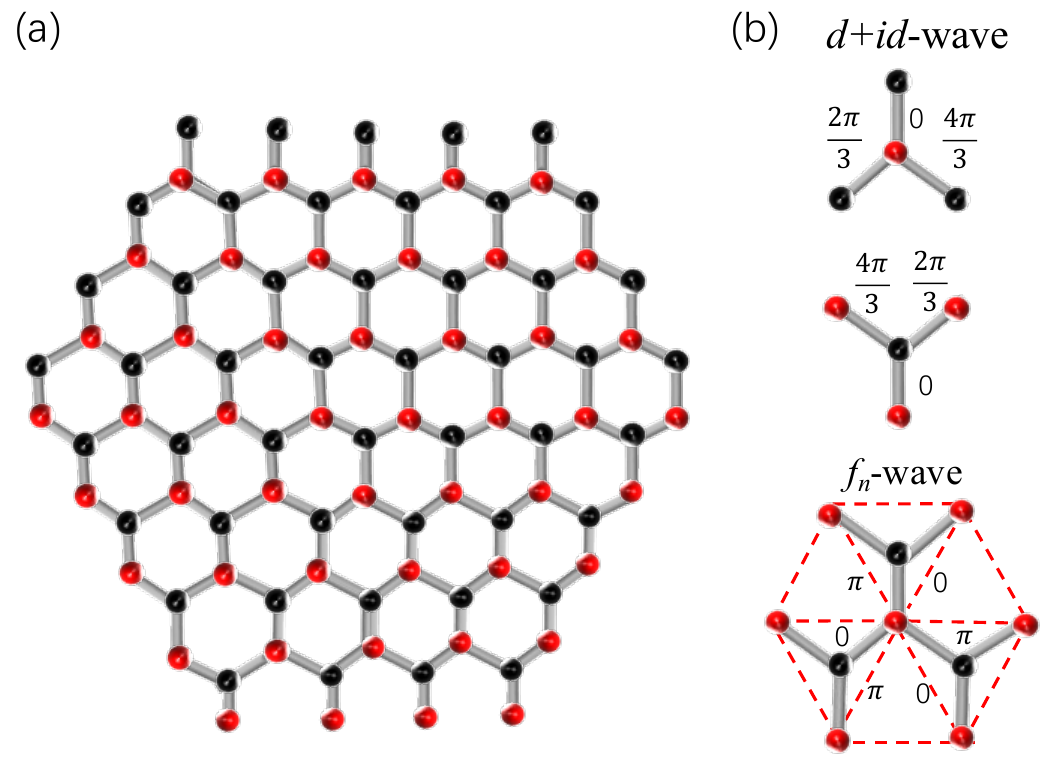}
\caption{\justifying (a) Sketch of graphene with double-48 sites; (b) Phases of the \(d+id\)-wave and \( f_n \)-wave pairing symmetries on the honeycomb lattice.}
\label{Fig1}
\vspace{0.2cm}
\end{figure}

In this study, we perform numerical calculations on honeycomb lattices containing double-48 and double-75 sites under periodic boundary conditions, using DQMC at finite temperature and CPMC at zero temperature, respectively. As depicted in Fig.~\ref{Fig1}(a), which corresponds to the double-48 lattice, the A sublattice is represented by red circles, while the B sublattice is indicated by black ones. Finite-temperature simulations are conducted using DQMC, in which the partition function is expressed as a high-dimensional integral over auxiliary fields and evaluated via Monte Carlo sampling. We employ the CPMC method to obtain ground-state properties. In CPMC, the ground-state wave function is projected from an initial wave function through a branching random walk within an overcomplete space of constrained Slater determinants, all of which maintain positive overlap with a chosen trial wave function. Extensive benchmarks have shown that the constrained-path approximation can yield reliable ground-state energies and correlation functions \cite{PhysRevB.55.7464}. However, recent benchmark studies have shown that the accuracy of CPMC results depends on the choice of trial wave function and that the systematic bias varies with the quality of the trial state \cite{PhysRevB.94.085103,PhysRevResearch.4.013239}. In our CPMC simulations, closed-shell electron fillings are used, adopting a constrained free-electron (CFE) trial wave function. As justified in detail in Appendix \ref{sec:A}, the observed pairing-channel trend is robust with respect to this CFE wave function by comparing the results obtained by a single-determinant generalized Hartree-Fock (GHF) trial wave function. To further test this choice, we have performed direct comparisons between our CPMC results and exact diagonalization (ED) on small clusters, the detailed benchmarking data are presented in Appendix \ref{sec:B}.

\color{black}Given that magnetic excitations may play a crucial role in mediating superconductivity in correlated electron systems, we introduce the zero-frequency spin susceptibility in the $z$-direction, defined as
\begin{align}
\chi(\mathbf{q}) = \int_{0}^{\beta} d\tau \sum_{d,d'=a,b} \sum_{i,j} e^{i\mathbf{q} \cdot (\mathbf{i_d} - \mathbf{j_{d'}})} \langle \mathbf{m}_{i_d}(\tau) \cdot \mathbf{m}_{j_{d'}}(0) \rangle,
\end{align}
where \(\mathbf{m}_{i_a}(\tau) = e^{H\tau} \mathbf{m}_{i_a}(0) e^{-H\tau}\) with the local moment operators defined as
\begin{align}
\mathbf{m}_{i_a} = a_{i\uparrow}^\dagger a_{i\uparrow} - a_{i\downarrow}^\dagger a_{i\downarrow}, \quad \mathbf{m}_{i_b} = b_{i\uparrow}^\dagger b_{i\uparrow} - b_{i\downarrow}^\dagger b_{i\downarrow}.
\end{align}
Within this formalism, the susceptibility \(\chi(\Gamma)\) characterizes ferromagnetic correlation, whereas \(\chi(M)\) probes antiferromagnetic correlation.

To characterize superconducting pairing tendencies in the honeycomb Hubbard model, we calculate the pairing susceptibilities for different symmetry channels:
\begin{align}
P_\alpha = \frac{1}{N_s} \sum_{i,j} \int_{0}^{\beta} d\tau \, \langle \Delta_\alpha^\dagger(i,\tau) \Delta_\alpha(j,0) \rangle .
\end{align}
The pairing correlation functions of different symmetry channels in CPMC are defined as
\begin{align}
C_\alpha(r) = \frac{1}{N} \sum_i \langle \Delta_\alpha^\dagger(\mathbf{r}_i + \mathbf{r}) \Delta_\alpha(\mathbf{r}_i) \rangle .
\end{align}
Here \(\alpha\) stands for the pairing symmetry. Owing to the on-site Hubbard interaction constraint in Eq.~(\ref{eq:model1}), inter-sublattice pairing is preferred, and the corresponding order parameter \(\Delta_\alpha^\dagger(i)\) takes the form \cite{PhysRevB.37.5070,PhysRevB.37.7359,PhysRevB.84.121410}
\begin{align}
\Delta_{\alpha}^{\dagger}(i) = \sum_{l} f_{\alpha}(\delta_l) \left( a_{i\uparrow} b_{i+\delta_l\downarrow} \pm a_{i\downarrow} b_{i+\delta_l\uparrow} \right)^{\dagger},
\label{eq6}
\end{align}
where \( f_{\alpha}(\delta_l) \) is the form factor of the pairing function. The vectors \( \delta_l \) correspond to the nearest-neighbor inter-sublattice or next-nearest-neighbor bonds illustrated in Fig.~\ref{Fig1}(b). The sign in Eq.~(\ref{eq6}) distinguishes spin-singlet \((-)\) and spin-triplet \((+)\) pairings. The form factors of main pairing symmetries are given by \cite{PhysRevB.77.235420}:
\begin{align}
f_{d+id}(\delta_l) = e^{i(l-1)\frac{2\pi}{3}}, \qquad l = 1, 2, 3,
\end{align}
\begin{align}
f_{f_n}(\delta_l) = e^{i \frac{1 + (-1)^{l}}{2} \pi}, \qquad l = 1, 2, 3, \ldots, 6.
\end{align}

To focus specifically on the effect of the interaction term on superconducting pairing symmetries, we recalculate the effective pairing susceptibility and the vertex contributions as follows:
\begin{align}
P_{\mathrm{eff}\,\alpha} = P_{\alpha} - \tilde{P}_{\alpha},
\end{align}
\begin{align}
V_{\alpha}(\mathbf{R}) = C_{\alpha}(\mathbf{R}) - \tilde{C}_{\alpha}(\mathbf{R}).
\end{align}
Here, $\tilde{P}_{\alpha}$ and $\tilde{C}_{\alpha}(\mathbf{R})$ represent the non-interacting single-particle contributions, obtained by replacing
$\langle a_{i\downarrow}^{\dagger}a_{j\downarrow}\;
b_{i+\delta_l\uparrow}^{\dagger}b_{j+\delta_{l'}\uparrow} \rangle$
\quad with \quad
$\langle a_{i\downarrow}^{\dagger}a_{j\downarrow} \rangle
\langle b_{i+\delta_l\uparrow}^{\dagger}b_{j+\delta_{l'}\uparrow} \rangle$.

\FloatBarrier 
\section{Results and Discussion}
\begin{figure}[!htbp]
\vspace{0.4cm}
\centering
\includegraphics[width=0.48\textwidth]{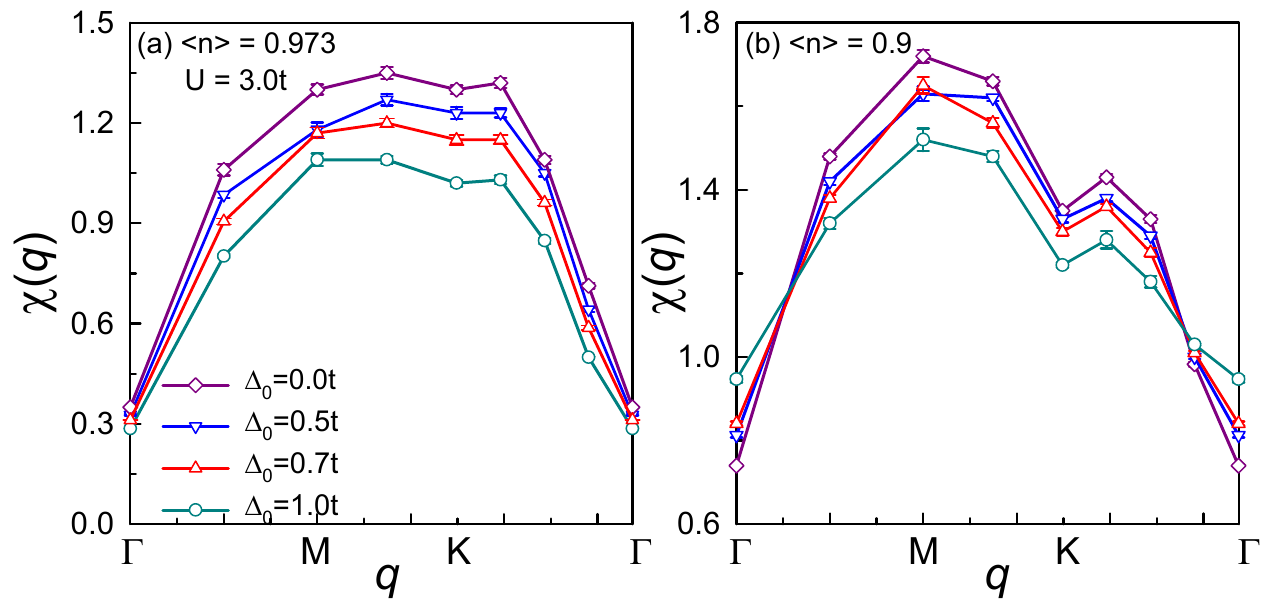}
\caption{\justifying The spin susceptibility $\chi(\mathbf{q})$ in q space for different staggered sublattice potential for (a) $\langle n\rangle = 0.973$; (b) $\langle n\rangle = 0.9$ with $T = t /6$ on a double-48 lattice.}
\label{fig:2}
\vspace{0.4cm}
\end{figure}

\vspace{0.3cm}
\label{sec:III}
In Fig.~\ref{fig:2}, we present the spin susceptibility $\chi(\mathbf{q})$ at a temperature of $T = t /6$ for two different electron fillings: $\langle n\rangle = 0.973$ and $\langle n\rangle = 0.9$. A comparison between Fig.~\ref{fig:2}(a) and (b) reveals distinct profiles of the spin susceptibility $\chi(\mathbf{q})$. In Fig.~\ref{fig:2}(a), the susceptibility at the $K$ and $M$ points is significantly larger than at the $\Gamma$ point, whereas in Fig.~\ref{fig:2}(b), a pronounced peak emerges at the $M$ point. These features indicate that antiferromagnetic spin correlations dominate at both electron concentrations. Furthermore, with increasing staggered sublattice potential $\Delta_0$, the overall magnitude of $\chi(\mathbf{q})$ is suppressed across momentum space, suggesting a progressive weakening of antiferromagnetic spin correlations under the applied potential at these doping levels. However, at a fixed electron density $\langle n \rangle = 0.9$, the susceptibility at the $\Gamma$ point increases with $\Delta_0$.

\begin{figure}[!htbp]
\centering
\includegraphics[width=0.48\textwidth]{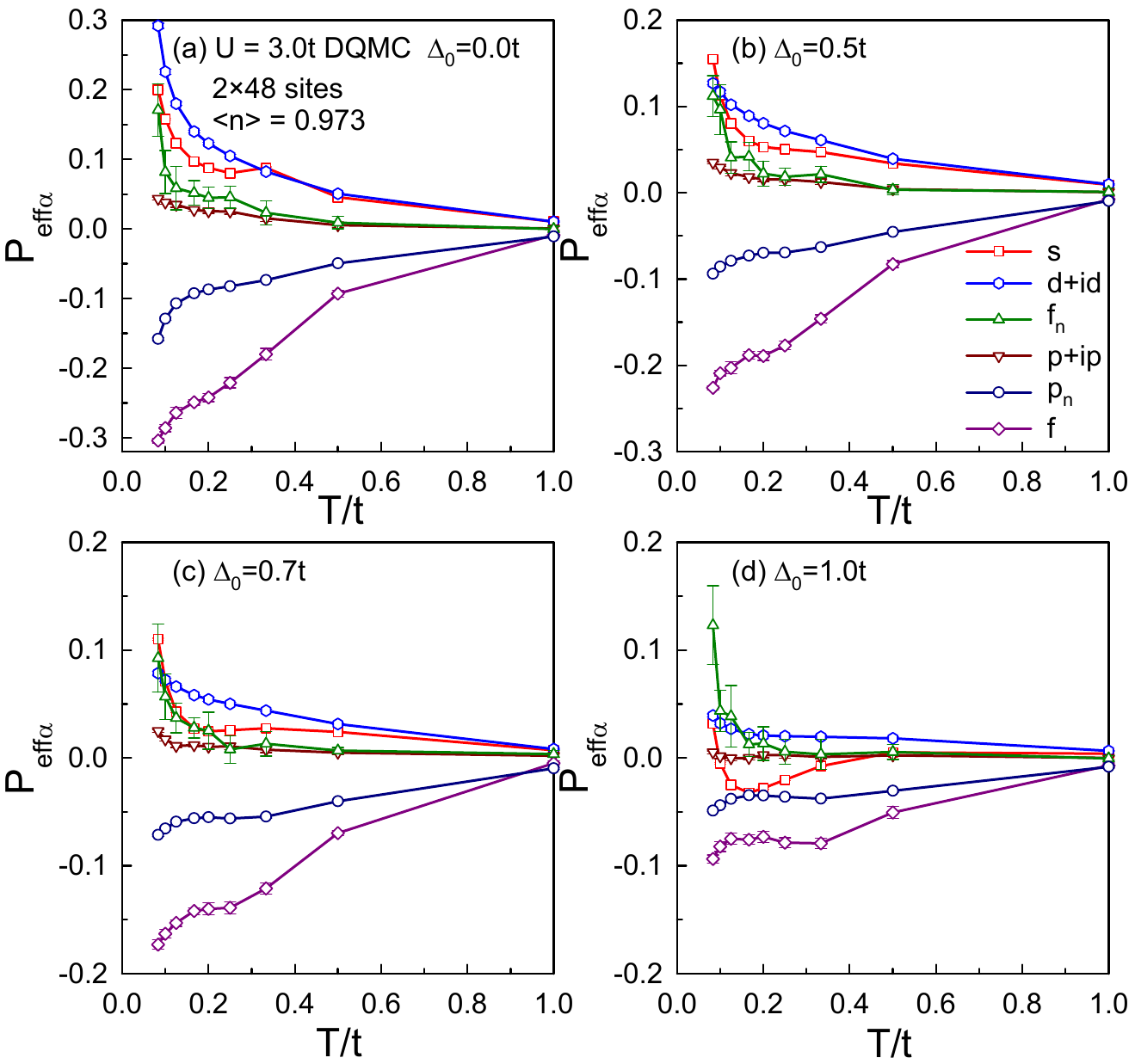}
\caption{\justifying Temperature dependence of the effective pairing susceptibilities for different pairing symmetries obtained by DQMC on a double-48 lattice at filling $\langle n\rangle = 0.973$, under various staggered sublattice potentials: (a) $\Delta_0 = 0.0t$; (b) $\Delta_0 = 0.5t$; (c) $\Delta_0 = 0.7t$; (d) $\Delta_0 = 1.0t$.}
\label{fig:3}
\vspace{0.1cm}
\end{figure}

To explore the superconducting pairing behavior, we examine the temperature dependence of the effective pairing susceptibilities for various pairing symmetries on the double-48 lattice at a filling of $\langle n\rangle = 0.973$ under  different staggered sublattice potentials by the DQMC method. As illustrated in Fig.~\ref{fig:3}, in the absence of a staggered potential, the $d+id$-wave pairing symmetry dominates, which is consistent with earlier studies of the honeycomb Hubbard model near half-filling \cite{PhysRevB.84.121410}. This behavior aligns with the conventional picture that antiferromagnetic spin fluctuations mediate $d+id$-wave pairing \cite{Scalapino1995TheCF}.

With increasing strength of the staggered sublattice potential, all pairing susceptibilities are generally suppressed. Meanwhile, upon lowering the temperature, the pairing susceptibility of the dominant pairing exhibits a significant increase, suggesting the possible emergence of the superconducting order \cite{PhysRevLett.62.1407,RevModPhys.84.1383}. In particular, when the staggered potential exceeds a critical value of $\Delta_c \approx 0.7t$, the $f_n$-wave pairing susceptibility surpasses that of the $d+id$-wave channel, establishing $f_n$-wave as the dominant pairing symmetry. This shift corresponds to a transition in the superconducting pairing symmetry.

While the preceding findings derive from the finite-temperature DQMC method, their extension to the low-temperature regime remains challenging due to the well-known fermion sign problem \cite{PhysRevB.41.9301}. To further examine whether long-range off-diagonal superconducting order emerges in the ground state, we employ the CPMC method. The CPMC controls the fermion sign problem through the constrained-path approximation \cite{PhysRevLett.74.3652}, wherein random walks are constrained to regions of positive overlap with the trial wave function, thereby enabling numerically stable projections to zero temperature and allowing a direct probe of possible superconducting pairing symmetries in the low-temperature limit.

\vspace{0.3cm}

\begin{figure}[!htbp]
\centering
\vspace{0.6cm}
\includegraphics[width=0.5\textwidth]{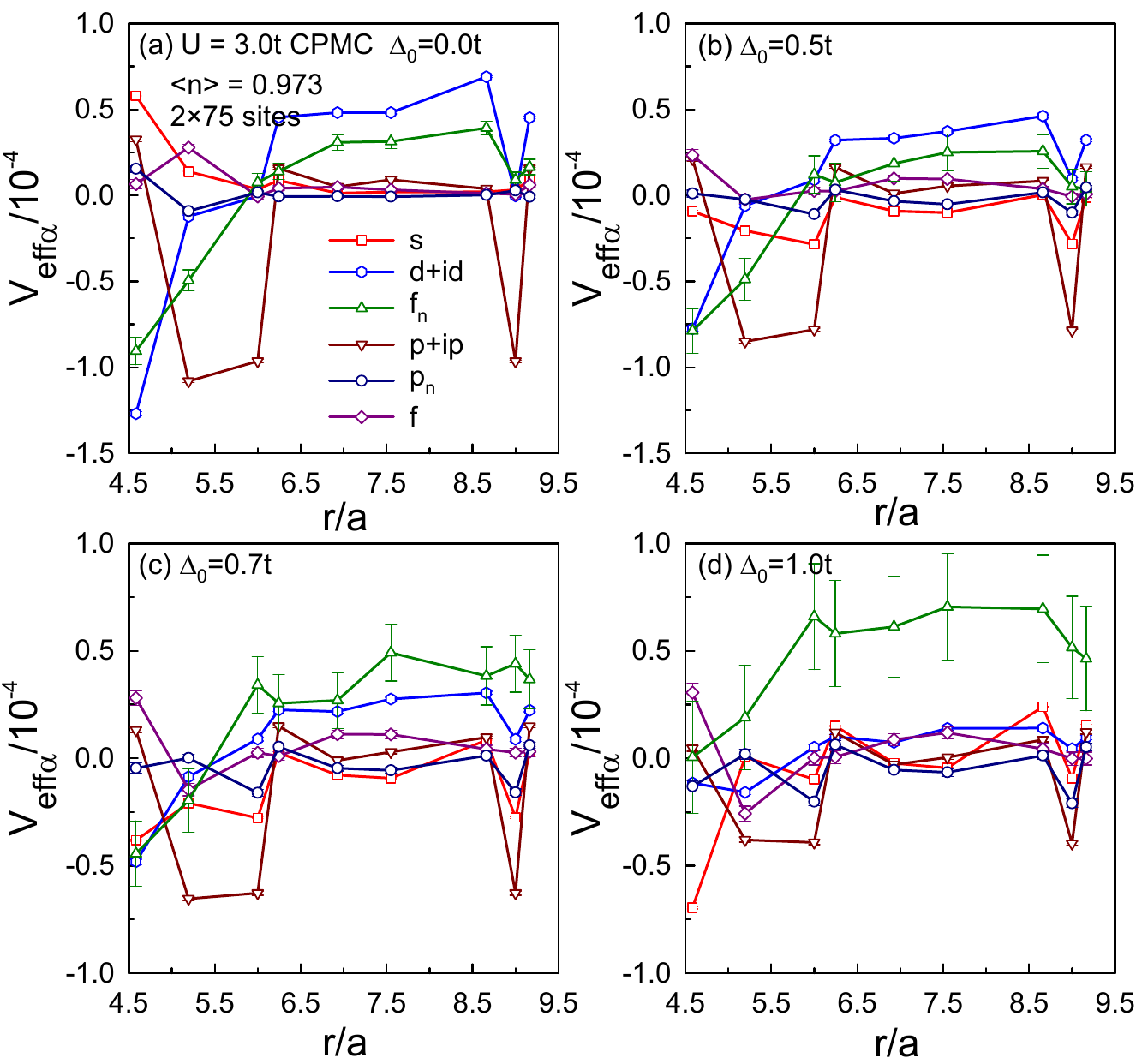}
\caption{\justifying Vertex contributions as a function of the normalized distance $r/a$ for different pairing symmetries obtained by CPMC on a double-75 lattice at filling $\langle n\rangle = 0.973$. Panels (a)--(d) correspond to different values of the staggered sublattice potential: (a) $\Delta_0 = 0.0t$; (b) $\Delta_0 = 0.5t$; (c) $\Delta_0 = 0.7t$; (d) $\Delta_0 = 1.0t$. Here $a$ denotes the lattice constant.}
\label{fig:4}
\end{figure}

\vspace{0.3cm}
\color{black} In Fig.~\ref{fig:4}, we show the long-range vertex contributions for different pairing symmetries calculated using the CPMC method on the double-75 lattice at the same filling $\langle n \rangle \approx 0.973$, corresponding to a closed-shell configuration. In the absence of a staggered potential, $d+id$-wave pairing clearly dominates. Upon introducing a finite sublattice potential, the $f_n$-wave vertex contribution is significantly enhanced, especially for long-range distances between electron pairs. Moreover, once the staggered potential exceeds a critical value of $\Delta_c \approx 0.7t$, the dominant superconducting pairing symmetry shifts from $d+id$-wave to $f_n$-wave, indicating a clear symmetry transition. This result further confirms that $f_n$-wave pairing prevails under a staggered potential in the lightly doped regime. These results are consistent with those obtained from DQMC calculations and underscore that tuning the A/B sublattice potential difference offers a viable route for steering the superconducting pairing symmetry in this class of materials. However, it should be noted that, due to the influence of the chosen trial wave function (as discussed in our Appendix \ref{sec:A}), the CPMC results show the same overall pairing symmetry trend across the tested trial wave functions, but the absolute magnitude of the \( f_n \)-wave connected vertex carries an additional systematic uncertainty originating from the trial wave function. 

\begin{figure}[!htbp]
\centering
\includegraphics[width=0.48\textwidth]{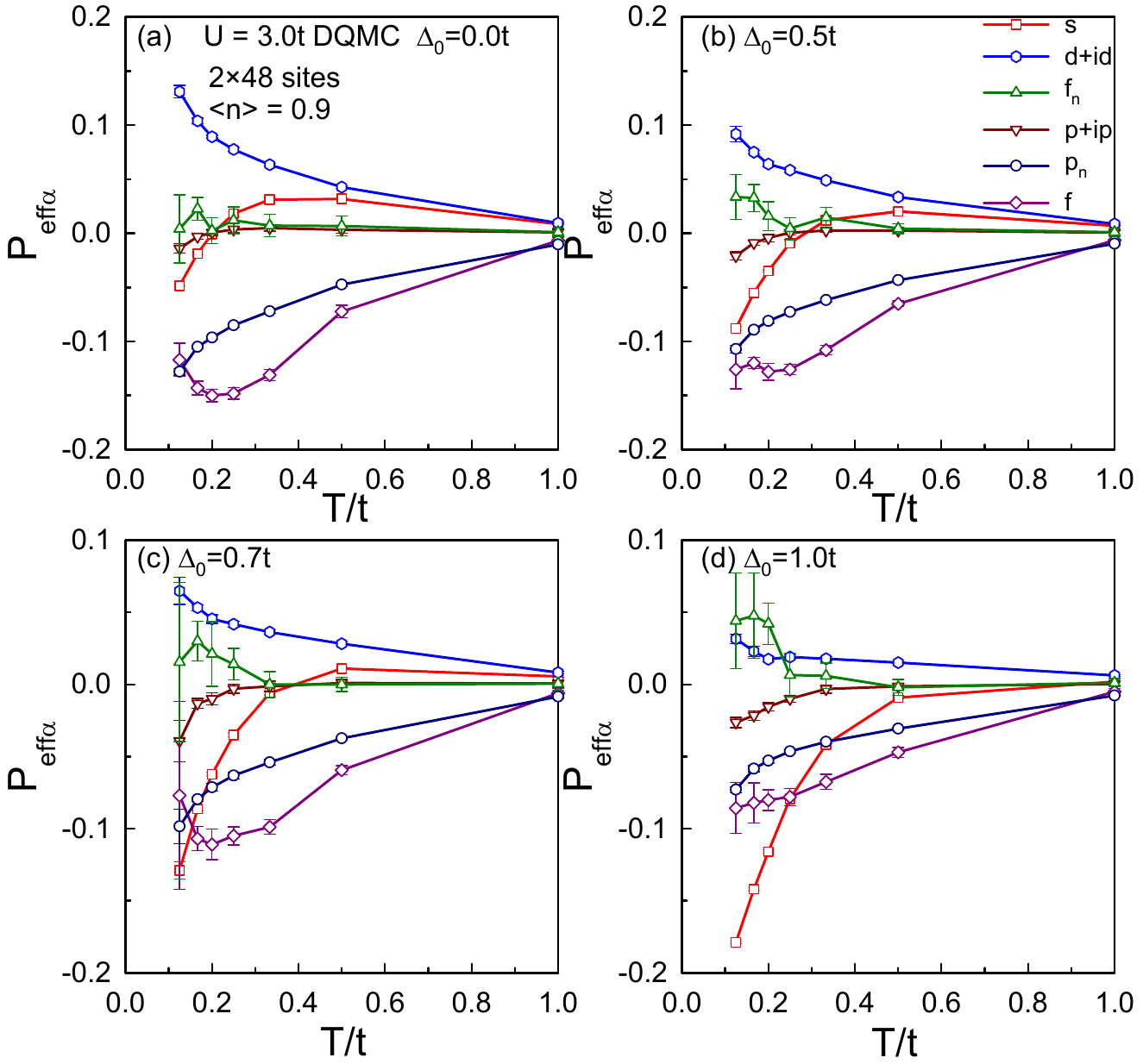}
\caption{\justifying Temperature dependence of the effective pairing susceptibilities for different pairing symmetries obtained by DQMC on a double-48 lattice at filling $\langle n\rangle = 0.9$, under various staggered sublattice potentials: (a) $\Delta_0 = 0.0t$; (b) $\Delta_0 = 0.5t$; (c) $\Delta_0 = 0.7t$;
(d) $\Delta_0 = 1.0t$.}
\label{fig:5}
\end{figure}

\begin{figure}[!htbp]
\vspace{0.3cm}
\centering
\includegraphics[width=0.48\textwidth]{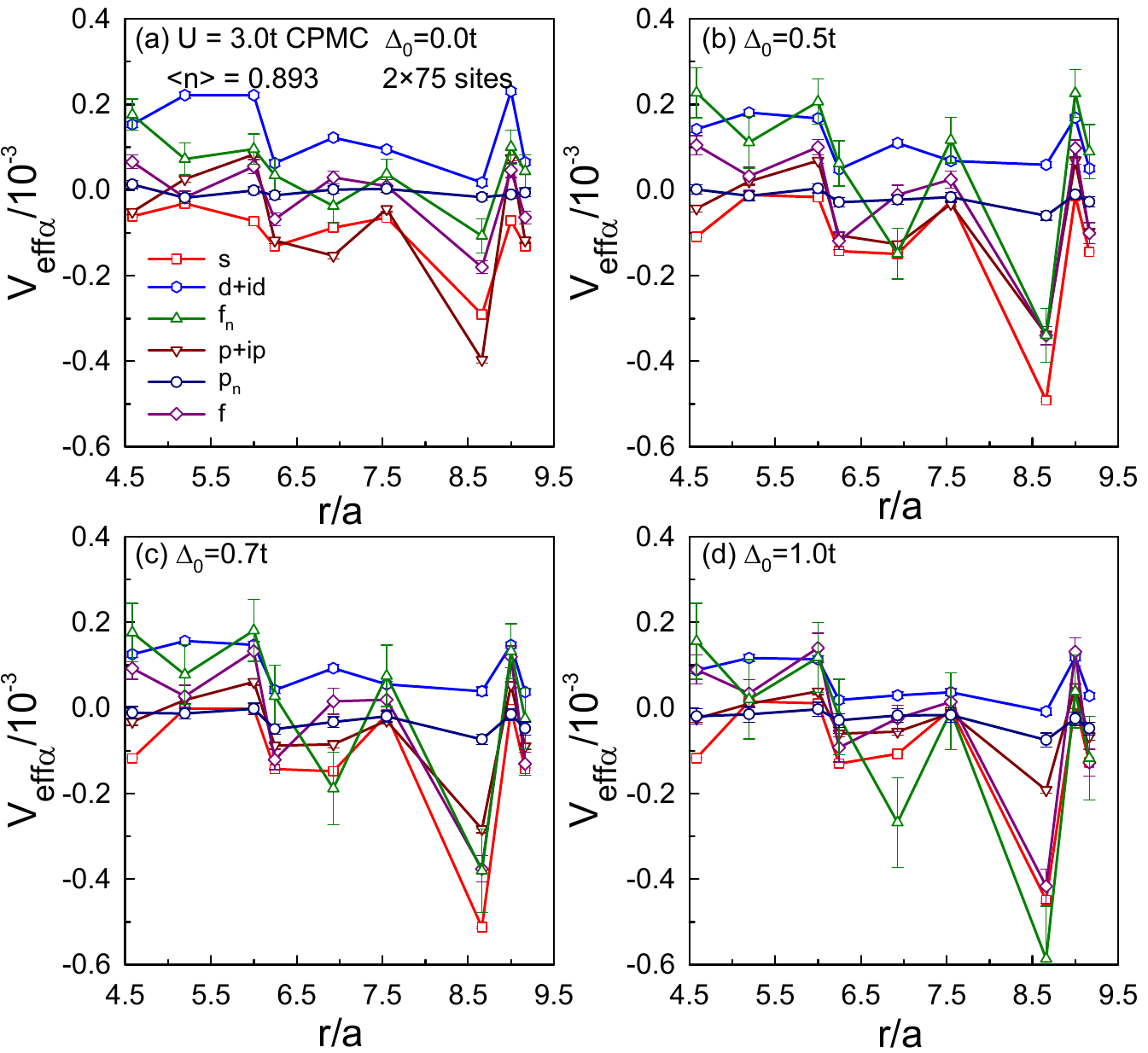}
\caption{\justifying Vertex contributions as a function of the normalized distance $r/a$ for different pairing symmetries obtained by CPMC on a double-75 lattice at filling $\langle n\rangle = 0.893$. Panels (a)--(d) correspond to different values of the staggered sublattice potential: (a) $\Delta_0 = 0.0t$; (b) $\Delta_0 = 0.5t$; (c) $\Delta_0 = 0.7t$; (d) $\Delta_0 = 1.0t$.}
\label{fig:6}
\vspace{0.3cm}
\end{figure}

\vspace{0.2cm}
To ensure generality, we also perform calculations at another electron concentration. The DQMC results, as shown in Fig.~\ref{fig:5}, indicate that as the doping concentration increases, the $f_n$-wave pairing is not significantly enhanced even in the presence of a stronger staggered potential, and the system remains dominated by $d+id$-wave pairing. This trend is consistently supported by the CPMC results presented in Fig.~\ref{fig:6}. This behavior may arise because the system is closer to the Van Hove singularity at this doping level, where repulsive interactions can be renormalized into an effective attraction, thereby stabilizing \(d+id\)-wave pairing as the dominant channel \cite{nandkishore_chiral_2012}.

\begin{figure}[!t]
\centering
\vspace{0.3cm}
\includegraphics[width=0.5\textwidth]{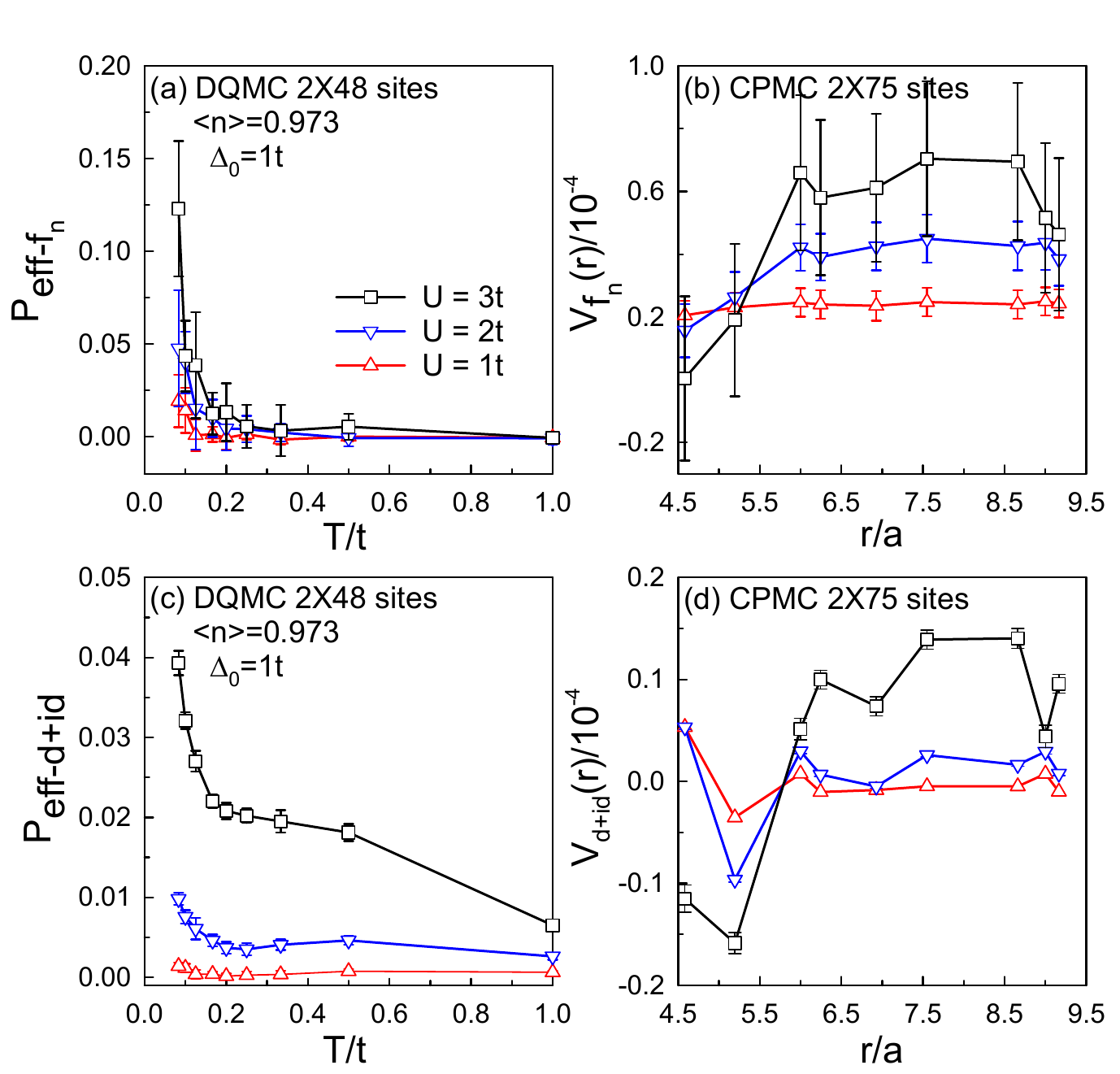}
\caption{\justifying Behavior of the effective pairing susceptibilities and vertex contributions for \( f_n \)-wave and $d+id$-wave symmetry under varying on-site interaction strengths \( U \). (a) Effective pairing susceptibilities for \( f_n \)-wave calculated by DQMC on a double-48 lattice at filling $\langle n\rangle = 0.973$ and $\Delta_0 = 1.0t$; (b) Vertex contributions for \( f_n \)-wave calculated by CPMC on a double-75 lattice at filling $\langle n\rangle = 0.973$ and $\Delta_0 = 1.0t$; (c) Effective pairing susceptibilities for \( d+id \)-wave calculated by DQMC at the same parameters; (d) Vertex contributions for \( d+id \)-wave calculated by CPMC.}
\label{fig:7}
\vspace{0.3cm}
\end{figure}

\vspace{0.2cm}
We then proceed to investigate the influence of the on-site interaction $U$ on the pairing symmetry for \( f_n \)-wave and $d+id$-wave under a fixed staggered potential of $\Delta_0 = 1.0t$ and an electron concentration of $\langle n\rangle = 0.973$. The results from both the DQMC and CPMC methods, presented in Fig.~\ref{fig:7}, demonstrate that irrespective of the computational approach, an increase in $U$ enhances the pairing tendencies in both pairing channels. Furthermore, the \( f_n \)-wave symmetry remains dominant throughout the entire $U$-range investigated, and no change in the leading pairing channel is observed within the $U$ range examined (as seen in Appendix \ref{sec:D}).

\vspace{0.2cm}
\color{black}The enhancement of the dominant $f_n$-wave pairing with increasing on-site interaction $U$ can be attributed to the specific electronic environment created by the staggered potential. In this regime near a band insulator, stronger $U$ amplifies interband electronic correlations, thereby promoting interband excitonic processes that selectively stabilize high-angular-momentum pairing channels such as the $f_n$-wave channel  \cite{doi:10.1073/pnas.2117735119}.

\section{Conclusion}
\label{sec:IV}
This study systematically investigates the effect of a staggered sublattice potential on the superconducting pairing symmetry in the doped honeycomb lattice Hubbard model, using determinant quantum Monte Carlo and constrained-path quantum Monte Carlo simulations. Our results reveal that at low doping levels, the introduction of a staggered potential suppresses antiferromagnetic fluctuations and drives a transition in the dominant pairing symmetry from $d+id$-wave to $f_n$-wave, whereas at high doping levels, the system remains dominated by $d+id$-wave pairing. Furthermore, an increase in the on-site interaction $U$ consistently enhances the dominant pairing channel, indicating that electronic correlations remain a key driver of superconducting pairing.

This work provides numerical results for understanding the tuning of unconventional superconducting symmetries in doped correlated electron systems. Notably, while prior approaches required substantial changes in doping concentration to alter pairing symmetry, our work demonstrates that tuning only the staggered potential is an effective pathway for controlling superconducting pairing symmetry. This finding provides inspiration for designing synthetic graphene-based superconductors with tailored pairing symmetries and may facilitate the exploration of unconventional and topological superconductivity in materials such as Li${}_x$MNCl and related systems.

\section{Acknowledgements}
\label{sec:V}
This work was supported by NSFC (12474218 and 12088101) and Beijing Natural Science
Foundation (No. 1242022 and 1252022). The numerical simulations in this work were performed at the HSCC of Beijing Normal University.

\section{DATA AVAILABILITY}
\label{sec:VI}
The data that support the findings of this article are openly available \cite{cai2026data}.

\newcounter{mytempeq}
\setcounter{mytempeq}{\value{equation}}

\appendix

\renewcommand{\theequation}{\arabic{equation}}

\vskip0.1in

\section{RELIABILITY ANALYSIS OF THE TRIAL WAVE FUNCTION SELECTED IN CPMC} 
\label{sec:A}
\renewcommand{\theequation}{\arabic{equation}}
\setcounter{equation}{\value{mytempeq}}

{ We employed the CPMC method to investigate the superconducting properties of the system by comparing the vertex contribution for different pairing symmetries. This method projects a trial wave function onto the ground state through a random walk in the space of Slater determinants via imaginary-time projection \cite{1sgg-ztw8}, which may introduce constrained-path bias. Constrained-path systematic errors are not universal constants. They depend on the trial wave function and can be more visible in correlation functions and connected vertex quantities than in the total energy \cite{PhysRevB.94.085103,PhysRevResearch.4.013239}. Therefore, it is necessary to examine the dependence of the pairing correlations on the choice of trial wave function.

In the original calculations we used a constrained free-electron (CFE) trial wave function. As an alternative, we constructed a single-determinant generalized Hartree-Fock (GHF) trial wave function. The GHF determinant was written in the spin-orbital basis,

\begin{align}
    |\Psi_T^{\mathrm{GHF}}\rangle =\prod_{m=1}^{N_e}\left(\sum_i u_{im}c^\dagger_{i\uparrow}+\sum_i v_{im} c^\dagger_{i\downarrow}\right)|0\rangle .
    \label{eq:ghf_trial}
\end{align}
\vspace{0.1cm}

Here each occupied orbital may contain both spin-up and spin-down components, which makes the trial state more flexible than a spin-separated free-electron determinant.
    
The GHF orbitals were obtained by solving a spin-orbital Hartree-Fock mean-field problem using the same one-body lattice matrix as in the corresponding CPMC production run. This one-body matrix contains the honeycomb hopping, periodic boundary condition, and staggered potential. For the on-site Hubbard interaction we used the standard local spin-density mean-field decoupling. Starting from the free-electron determinant, we performed several independent GHF mean-field calculations and selected the determinant with the lowest Hartree-Fock total energy as the alternative CPMC trial wave function.
    
We then compared the back-propagated connected pairing vertex obtained with the CFE trial and the GHF trial. The calculations used the same cluster and filling as in the manuscript, with \(L_x=L_y=5\), \(N_s=150\), \(N_\uparrow=N_\downarrow=73\) and \(U=3t\) under periodic boundary conditions. The same radial distance grouping as in the original pairing-correlation data was used. 

\begin{figure}[!h]
    \vspace{0.1cm}
	\includegraphics[width=0.46\textwidth]{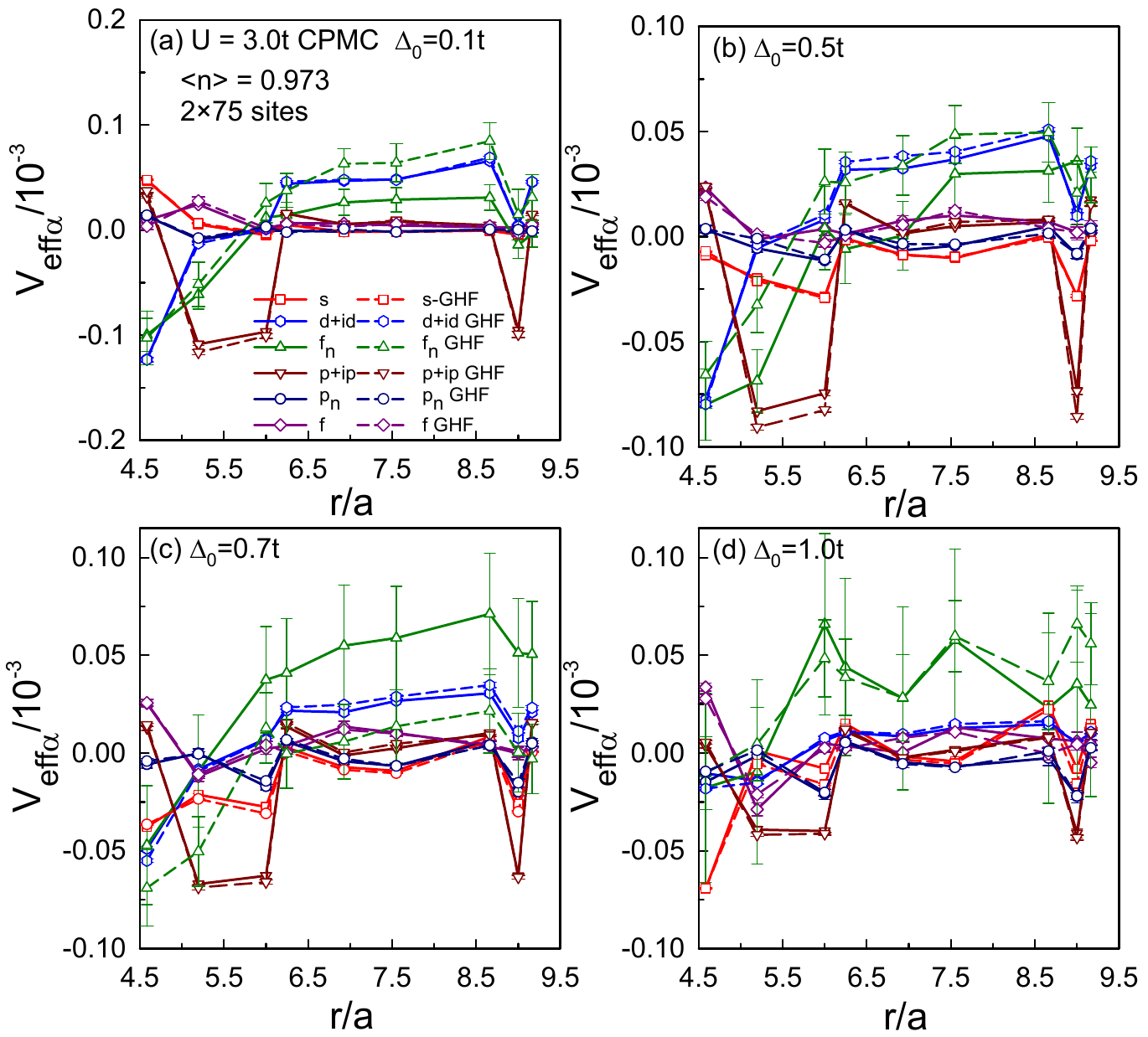}
	\caption{\justifying \label{FigA1} Trial-wave-function dependence of the CPMC connected pairing vertex. The original CFE trial wave function is compared with a single-determinant GHF trial wave function.}
    \vspace{0.1cm}
    \end{figure}

As shown in Fig.~\ref{FigA1}, the main pairing-channel trends are stable under this change of trial wave function. In particular, the \(d+id\)-wave vertex is strongly suppressed as the staggered potential is increased. The \(p_n\)-wave channel remains small, and the \(s\)-wave channel shows the same trend for the two trial wave functions. Therefore, the main conclusion, namely that the staggered potential tunes the competition among pairing channels, is not a consequence of using only the CFE trial.
    
We also find that the connected \(f_n\)-wave vertex is more sensitive to the trial wave function, especially in its absolute magnitude. This sensitivity is plausible because the \(f_n\)-wave form factor contains sign-changing bond combinations, and the connected vertex is obtained after subtracting the disconnected bubble contribution from the full pairing correlation. Small trial-dependent changes in the one-body density matrix can therefore be amplified in this connected quantity. For this reason, although the qualitative trend is stable, the absolute magnitude of the \(f_n\)-wave vertex should not be regarded as a trial-independent quantity.

\setcounter{mytempeq}{\value{equation}}

\color{black}
\section{BENCHMARKING OF CPMC ON SMALL CLUSTERS WITH EXACT DIAGONALIZATION} 
\label{sec:B}
\vspace*{-6pt}
\renewcommand{\theequation}{\arabic{equation}}
\setcounter{equation}{\value{mytempeq}}

{ Recognizing that the reliability of CPMC results depends on the quality of the trial wave function, we have performed a small-cluster check using the ED on a \(2\times3\times3\) honeycomb lattice cluster, with the identical Hubbard parameters employed in our production runs at the staggered potential \(\Delta_0=0.7t\). }{In CPMC calculations, the ground-state energy can be evaluated using the mixed estimator. However, for non-commuting observables, such as pairing correlations, vertex contributions, spin correlations, charge correlations and double occupancy, the mixed estimator generally introduces a bias, and back propagation is required for their evaluation. Therefore, in our CPMC calculations, all these observables are evaluated using the back-propagation method.} We compared the ED benchmark data with our CPMC back-propagated estimates for both the ground-state energies and the short-range correlation functions.

In Table~\ref{tab:h18-ed-cpmc-short-range}, we showed a comparison of the CPMC method with the ED method on the \(2\times3\times3\) honeycomb lattice for the energy, double occupancy, nearest-neighbor spin correlation and nearest-neighbor charge correlation. To avoid mixing this benchmark with artificial shell effects, we reported results for the closed-shell \(N_\uparrow=N_\downarrow=7\) sector under periodic boundary conditions. 

The local observables are
\setlength{\jot}{1pt}
\begin{equation}
    \begin{gathered}
    D=\frac{1}{N_s}\sum_i\langle n_{i\uparrow}n_{i\downarrow}\rangle,
    \quad
    S^z_{\rm nn}=\frac{1}{N_b}\sum_{\langle ij\rangle}\langle S_i^zS_j^z\rangle,
    \\[1mm]
    C_{\rm nn}=\frac{1}{N_b}\sum_{\langle ij\rangle}\langle n_i n_j\rangle .
    \end{gathered}
    \label{E}
\end{equation}
\setlength{\jot}{1pt}
Here, $N_s$ denotes the number of sublattice sites and $N_b$ is the number of nearest-neighbor bonds. $n_i$ is the number operator at site \( \mathbf{R}_i \).

     \begin{table}[htbp]
     
     \centering
     \caption{\justifying Comparison of the ED and the CPMC results on a \(2\times3\times3\) honeycomb lattice cluster at \(\Delta_0=0.7t\), \(N_\uparrow=N_\downarrow=7\) under periodic boundary conditions. The CPMC rows are back-propagated estimates averaged over independent replicas with 512 walkers per replica: 16 replicas for \(U=1t, 2t, 3t\) and 64 replicas for \(U=4t\). Parentheses denote the standard error in the last quoted digit, rounded to one significant digit.}
     
    \begin{tabular}{cccccc}
    \toprule
    \(U\) & Method & \(E_T\) & D & \(S^z_{\rm nn}\) & \(C_{\rm nn}\) \\

    \hline  
    \(1.0\) & ED   & -25.757    & 0.14399    & -0.03153    & 0.45099 \\
            & CPMC & -25.755(3) & 0.14406(2) & -0.03151(1) & 0.45090(3) \\
            \hline  
    \(2.0\) & ED   & -23.379    & 0.12059    & -0.03613    & 0.47525 \\
            & CPMC & -23.373(6) & 0.12078(4) & -0.03609(3) & 0.47503(7) \\
            \hline  
    \(3.0\) & ED   & -21.40     & 0.09957    & -0.04064    & 0.4950 \\
            & CPMC & -21.40(1)  & 0.09972(7) & -0.04066(5) & 0.4949(1) \\
             \hline  
    \(4.0\) & ED   & -19.779    & 0.08114    & -0.04482    & 0.51101 \\
            & CPMC & -19.759(9) & 0.08145(5) & -0.04466(4) & 0.5106(1) \\

    \hline
	\hline  
    \end{tabular}
    \label{tab:h18-ed-cpmc-short-range}
    \end{table}

As shown in Table~\ref{tab:h18-ed-cpmc-short-range}, the CPMC energies agree well with the ED values on this closed-shell 18-site benchmark. For \(U=1t, 2t, 3t\), the ED-CPMC energy differences are within the quoted one-standard-error CPMC statistical error bars. For \(U=4t\), the remaining energy difference is about \(2\times 10^{-2}t\), corresponding to roughly two conservative standard errors. The local short-range quantities show similarly small absolute deviations. The double occupancy, nearest-neighbor spin correlation, and nearest-neighbor charge correlation agree with ED at the \(10^{-4}\)--\(10^{-3}\) level throughout the table. Although some of these local-observable deviations are larger than the quoted Monte Carlo standard errors, their absolute size is small on the scale relevant for the short-range correlation analysis. This benchmark therefore supports the conclusion that the CPMC back-propagated estimator does not produce a qualitative discrepancy in the local quantities entering the pairing analysis.
}
\vspace*{-6pt}

\setcounter{mytempeq}{\value{equation}}

\section{CPMC RESULTS ON OTHER SYSTEM SIZES} 
\label{sec:C}
\vspace*{-6pt}
\renewcommand{\theequation}{\arabic{equation}}
\setcounter{equation}{\value{mytempeq}}

{\color{black}
Since pairing correlations in the doped Hubbard model can be highly sensitive to system size and cluster geometry \cite{xu_coexistence_2024}, we performed additional large-scale simulations for the system size $L=6$ in representative parameter sets (staggered potential $\Delta_0$ = 0$t$, 0.5$t$, 0.7$t$, 1$t$  with $U=3t$ ) using the CPMC method.

 As shown in Fig.~\ref{FigA2} and Fig.~\ref{FigA3}, the results exhibit clear size-consistent behavior. In the small-doping regime, at the filling $\langle n\rangle=0.981$, corresponding to a closed-shell configuration, $d+id$-wave pairing is the dominant pairing channel without a staggered potential. The application of staggered potential drives a transition from $d+id$-wave to \( f_n \)-wave pairing, while at large doping for $\langle n\rangle=0.926$, the $d+id$-wave channel remains dominant. {These findings agree with the results of $L=5$, providing further support for the robustness of the qualitative pairing-channel trend. However, since these calculations still employ periodic boundary conditions and correspond to slightly different closed-shell fillings, they should not be viewed as a complete assessment of finite-size or boundary-condition effects. A full thermodynamic-limit extrapolation or twist-averaged boundary-condition analysis is beyond the scope of the present work.}

	\begin{figure}[!h]
    \vspace{0.1cm}
	\includegraphics[width=0.47\textwidth,height=6.95cm]{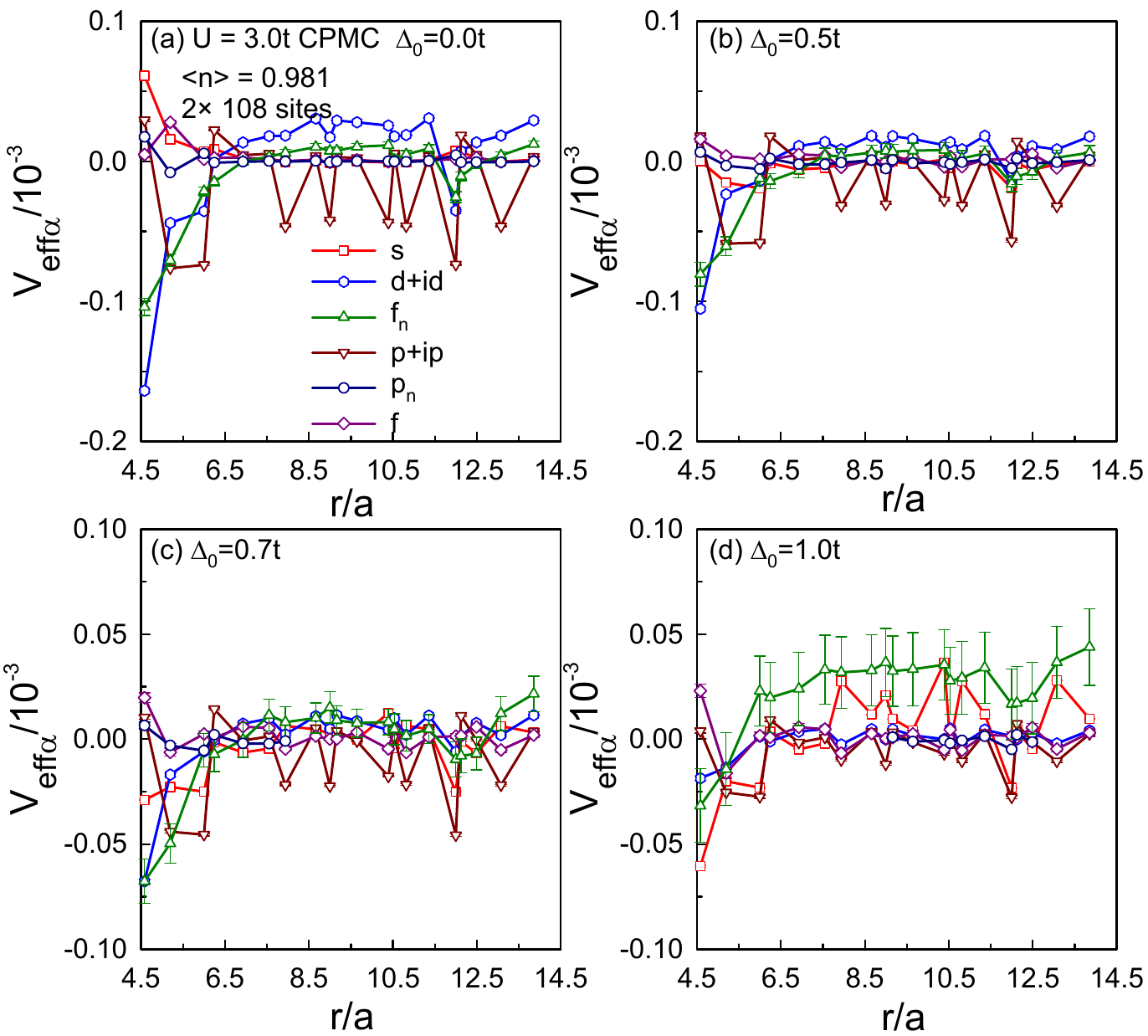}
	\caption{\justifying \label{FigA2} Vertex contributions for different pairing symmetries obtained by CPMC on a double-108 lattice at filling $\langle n\rangle = 0.981$.}
    \vspace{0.1cm}
    \end{figure}

\begin{figure}[h!]
	\includegraphics[width=0.47\textwidth]{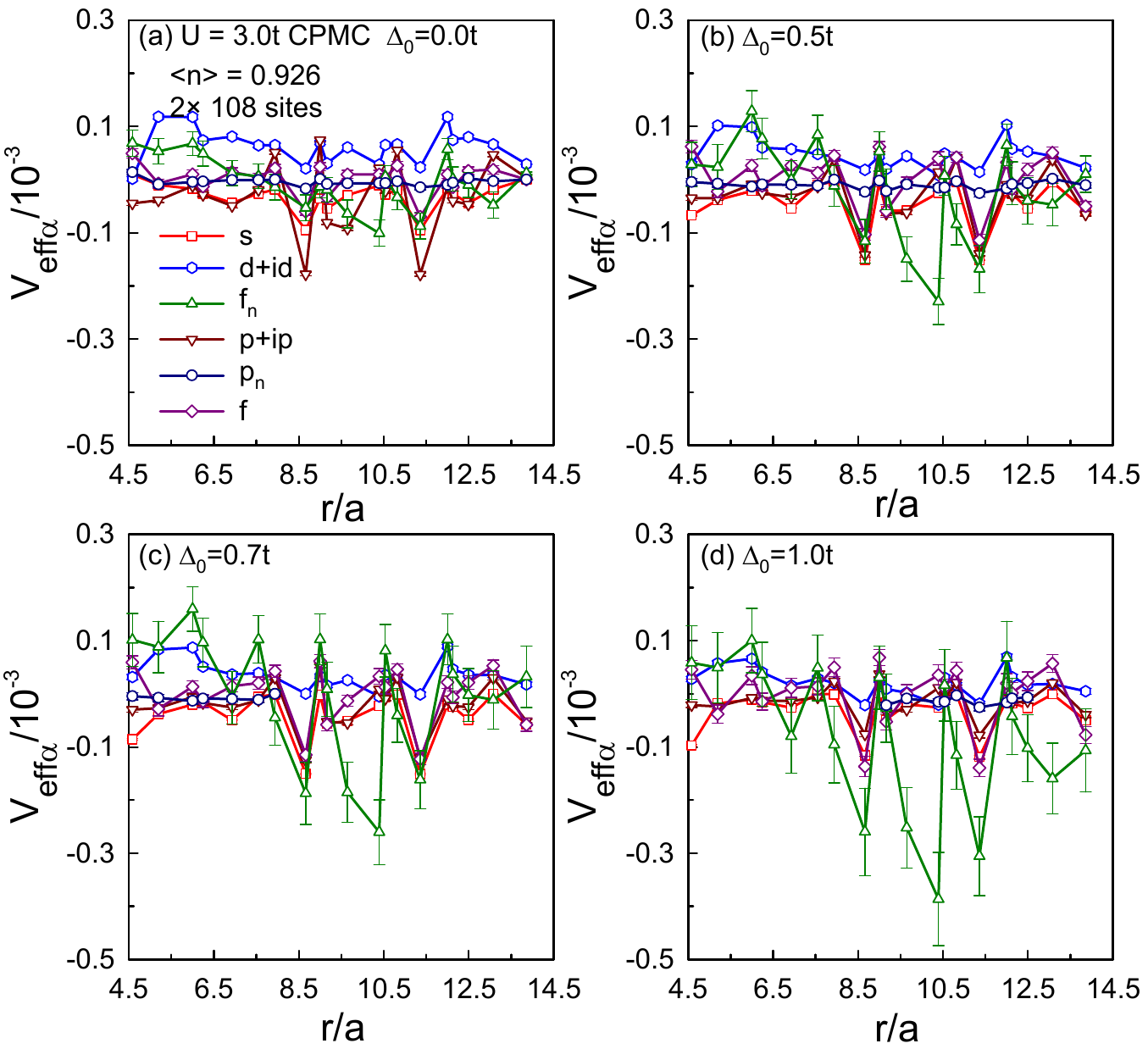}
	\caption{\justifying \label{FigA3} Vertex contributions for different pairing symmetries obtained by CPMC on a double-108 lattice at filling $\langle n\rangle = 0.926$.}
\end{figure}

}

\setcounter{mytempeq}{\value{equation}}
\vskip0.1in
\section{U-DEPENDENCE ANALYSIS} 
\label{sec:D}

\renewcommand{\theequation}{\arabic{equation}}
\setcounter{equation}{\value{mytempeq}}

{In order to further test the robustness of our conclusion, we have calculated the long-range-averaged vertex contribution for each pairing symmetry as a function of $U$. The long-range-averaged vertex contributions take the form \cite{PhysRevB.101.155413}

\begin{equation}
    \overline{V}_{eff\alpha} = \frac{1}{\sqrt{N'}} \sum_{r > 4a} V_{eff\alpha}(r),
\end{equation}
where \(N'\) is the number of electron pairs with \(r > 4a\).

As shown in Fig.~\ref{FigA4}, the long-range-averaged vertex
contributions \(\overline{V}_{eff\alpha}\) for different pairing symmetries vary with on-site interaction strength \( U \). We observed that long-range-averaged vertex contributions \(\overline{V}_{eff\alpha}\) for $d+id$ and $f_n$-wave symmetries increase with $U$. In addition, the long-range-averaged vertex contribution of the \( f_n \)-wave symmetry remains markedly larger than other pairing channels across the entire parameter range, confirming the dominance of \( f_n \)-wave pairing.

\begin{figure}[!h]
	\includegraphics[width=0.4\textwidth,height=0.34\textwidth]{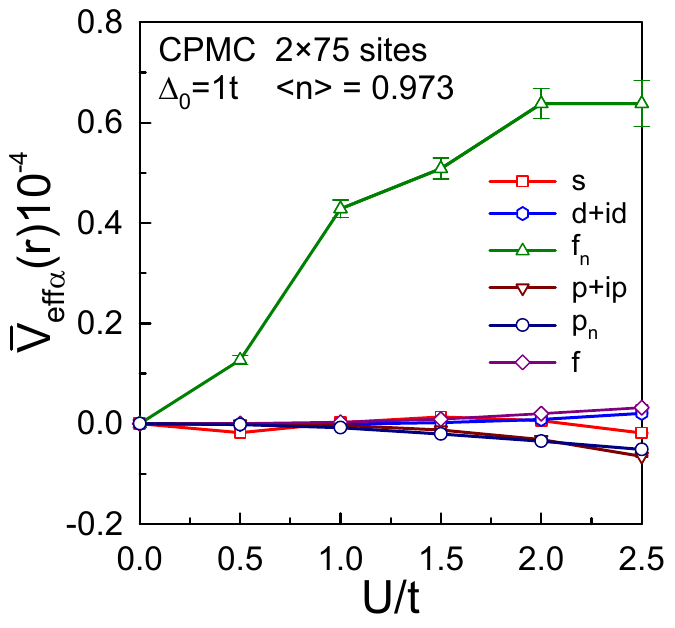}
	\caption{\justifying \label{FigA4} Behavior of the long-range-averaged vertex contributions for different pairing symmetries under varying on-site interaction strengths \( U \), calculated by CPMC on a double-75 lattice at filling $\langle n\rangle = 0.973$ and $\Delta_0 = 1.0t$.}
    \vspace{0.3cm}
    \end{figure}

}

\bibliographystyle{apsrev4-2}
\bibliography{SPHMReferences}

\begin{thebibliography}{58}%
\makeatletter
\providecommand \@ifxundefined [1]{%
 \@ifx{#1\undefined}
}%
\providecommand \@ifnum [1]{%
 \ifnum #1\expandafter \@firstoftwo
 \else \expandafter \@secondoftwo
 \fi
}%
\providecommand \@ifx [1]{%
 \ifx #1\expandafter \@firstoftwo
 \else \expandafter \@secondoftwo
 \fi
}%
\providecommand \natexlab [1]{#1}%
\providecommand \enquote  [1]{``#1''}%
\providecommand \bibnamefont  [1]{#1}%
\providecommand \bibfnamefont [1]{#1}%
\providecommand \citenamefont [1]{#1}%
\providecommand \href@noop [0]{\@secondoftwo}%
\providecommand \href [0]{\begingroup \@sanitize@url \@href}%
\providecommand \@href[1]{\@@startlink{#1}\@@href}%
\providecommand \@@href[1]{\endgroup#1\@@endlink}%
\providecommand \@sanitize@url [0]{\catcode `\\12\catcode `\$12\catcode
  `\&12\catcode `\#12\catcode `\^12\catcode `\_12\catcode `\%12\relax}%
\providecommand \@@startlink[1]{}%
\providecommand \@@endlink[0]{}%
\providecommand \url  [0]{\begingroup\@sanitize@url \@url }%
\providecommand \@url [1]{\endgroup\@href {#1}{\urlprefix }}%
\providecommand \urlprefix  [0]{URL }%
\providecommand \Eprint [0]{\href }%
\providecommand \doibase [0]{https://doi.org/}%
\providecommand \selectlanguage [0]{\@gobble}%
\providecommand \bibinfo  [0]{\@secondoftwo}%
\providecommand \bibfield  [0]{\@secondoftwo}%
\providecommand \translation [1]{[#1]}%
\providecommand \BibitemOpen [0]{}%
\providecommand \bibitemStop [0]{}%
\providecommand \bibitemNoStop [0]{.\EOS\space}%
\providecommand \EOS [0]{\spacefactor3000\relax}%
\providecommand \BibitemShut  [1]{\csname bibitem#1\endcsname}%
\let\auto@bib@innerbib\@empty
\bibitem [{\citenamefont {Read}\ and\ \citenamefont
  {Green}(2000)}]{PhysRevB.61.10267}%
  \BibitemOpen
  \bibfield  {author} {\bibinfo {author} {\bibfnamefont {N.}~\bibnamefont
  {Read}}\ and\ \bibinfo {author} {\bibfnamefont {D.}~\bibnamefont {Green}},\
  }\href {https://doi.org/10.1103/PhysRevB.61.10267} {\bibfield  {journal}
  {\bibinfo  {journal} {Phys. Rev. B}\ }\textbf {\bibinfo {volume} {61}},\
  \bibinfo {pages} {10267} (\bibinfo {year} {2000})}\BibitemShut {NoStop}%
\bibitem [{\citenamefont {Zheng}(2023)}]{zheng_high_2023}%
  \BibitemOpen
  \bibfield  {author} {\bibinfo {author} {\bibfnamefont {G.-q.}\ \bibnamefont
  {Zheng}},\ }\href {https://doi.org/10.1088/1742-6596/2545/1/012001}
  {\bibfield  {journal} {\bibinfo  {journal} {Journal of Physics: Conference
  Series}\ }\textbf {\bibinfo {volume} {2545}},\ \bibinfo {pages} {012001}
  (\bibinfo {year} {2023})}\BibitemShut {NoStop}%
\bibitem [{\citenamefont {Mackenzie}\ and\ \citenamefont
  {Maeno}(2003)}]{RevModPhys.75.657}%
  \BibitemOpen
  \bibfield  {author} {\bibinfo {author} {\bibfnamefont {A.~P.}\ \bibnamefont
  {Mackenzie}}\ and\ \bibinfo {author} {\bibfnamefont {Y.}~\bibnamefont
  {Maeno}},\ }\href {https://doi.org/10.1103/RevModPhys.75.657} {\bibfield
  {journal} {\bibinfo  {journal} {Rev. Mod. Phys.}\ }\textbf {\bibinfo {volume}
  {75}},\ \bibinfo {pages} {657} (\bibinfo {year} {2003})}\BibitemShut
  {NoStop}%
\bibitem [{\citenamefont {Ran}\ \emph {et~al.}(2019)\citenamefont {Ran},
  \citenamefont {Eckberg}, \citenamefont {Ding}, \citenamefont {Furukawa},
  \citenamefont {Metz}, \citenamefont {Saha}, \citenamefont {Liu},
  \citenamefont {Zic}, \citenamefont {Kim}, \citenamefont {Paglione},\ and\
  \citenamefont {Butch}}]{doi:10.1126/science.aav8645}%
  \BibitemOpen
  \bibfield  {author} {\bibinfo {author} {\bibfnamefont {S.}~\bibnamefont
  {Ran}}, \bibinfo {author} {\bibfnamefont {C.}~\bibnamefont {Eckberg}},
  \bibinfo {author} {\bibfnamefont {Q.-P.}\ \bibnamefont {Ding}}, \bibinfo
  {author} {\bibfnamefont {Y.}~\bibnamefont {Furukawa}}, \bibinfo {author}
  {\bibfnamefont {T.}~\bibnamefont {Metz}}, \bibinfo {author} {\bibfnamefont
  {S.~R.}\ \bibnamefont {Saha}}, \bibinfo {author} {\bibfnamefont {I.-L.}\
  \bibnamefont {Liu}}, \bibinfo {author} {\bibfnamefont {M.}~\bibnamefont
  {Zic}}, \bibinfo {author} {\bibfnamefont {H.}~\bibnamefont {Kim}}, \bibinfo
  {author} {\bibfnamefont {J.}~\bibnamefont {Paglione}},\ and\ \bibinfo
  {author} {\bibfnamefont {N.~P.}\ \bibnamefont {Butch}},\ }\href
  {https://doi.org/10.1126/science.aav8645} {\bibfield  {journal} {\bibinfo
  {journal} {Science}\ }\textbf {\bibinfo {volume} {365}},\ \bibinfo {pages}
  {684} (\bibinfo {year} {2019})}\BibitemShut {NoStop}%
\bibitem [{\citenamefont {Contamin}\ \emph {et~al.}(2021)\citenamefont
  {Contamin}, \citenamefont {Delbecq}, \citenamefont {Douçot}, \citenamefont
  {Cottet},\ and\ \citenamefont {Kontos}}]{contamin_hybrid_2021}%
  \BibitemOpen
  \bibfield  {author} {\bibinfo {author} {\bibfnamefont {L.~C.}\ \bibnamefont
  {Contamin}}, \bibinfo {author} {\bibfnamefont {M.~R.}\ \bibnamefont
  {Delbecq}}, \bibinfo {author} {\bibfnamefont {B.}~\bibnamefont {Douçot}},
  \bibinfo {author} {\bibfnamefont {A.}~\bibnamefont {Cottet}},\ and\ \bibinfo
  {author} {\bibfnamefont {T.}~\bibnamefont {Kontos}},\ }\href
  {https://doi.org/10.1038/s41534-021-00508-w} {\bibfield  {journal} {\bibinfo
  {journal} {npj Quantum Information}\ }\textbf {\bibinfo {volume} {7}},\
  \bibinfo {pages} {171} (\bibinfo {year} {2021})}\BibitemShut {NoStop}%
\bibitem [{\citenamefont {Gastiasoro}\ \emph {et~al.}(2020)\citenamefont
  {Gastiasoro}, \citenamefont {Ruhman},\ and\ \citenamefont
  {Fernandes}}]{GASTIASORO2020168107}%
  \BibitemOpen
  \bibfield  {author} {\bibinfo {author} {\bibfnamefont {M.~N.}\ \bibnamefont
  {Gastiasoro}}, \bibinfo {author} {\bibfnamefont {J.}~\bibnamefont {Ruhman}},\
  and\ \bibinfo {author} {\bibfnamefont {R.~M.}\ \bibnamefont {Fernandes}},\
  }\href {https://doi.org/https://doi.org/10.1016/j.aop.2020.168107} {\bibfield
   {journal} {\bibinfo  {journal} {Annals of Physics}\ }\textbf {\bibinfo
  {volume} {417}},\ \bibinfo {pages} {168107} (\bibinfo {year}
  {2020})}\BibitemShut {NoStop}%
\bibitem [{\citenamefont {Taguchi}\ \emph {et~al.}(2006)\citenamefont
  {Taguchi}, \citenamefont {Kitora},\ and\ \citenamefont
  {Iwasa}}]{PhysRevLett.97.107001}%
  \BibitemOpen
  \bibfield  {author} {\bibinfo {author} {\bibfnamefont {Y.}~\bibnamefont
  {Taguchi}}, \bibinfo {author} {\bibfnamefont {A.}~\bibnamefont {Kitora}},\
  and\ \bibinfo {author} {\bibfnamefont {Y.}~\bibnamefont {Iwasa}},\ }\href
  {https://doi.org/10.1103/PhysRevLett.97.107001} {\bibfield  {journal}
  {\bibinfo  {journal} {Phys. Rev. Lett.}\ }\textbf {\bibinfo {volume} {97}},\
  \bibinfo {pages} {107001} (\bibinfo {year} {2006})}\BibitemShut {NoStop}%
\bibitem [{\citenamefont {Kasahara}\ \emph {et~al.}(2009)\citenamefont
  {Kasahara}, \citenamefont {Kishiume}, \citenamefont {Takano}, \citenamefont
  {Kobayashi}, \citenamefont {Matsuoka}, \citenamefont {Onodera}, \citenamefont
  {Kuroki}, \citenamefont {Taguchi},\ and\ \citenamefont
  {Iwasa}}]{PhysRevLett.103.077004}%
  \BibitemOpen
  \bibfield  {author} {\bibinfo {author} {\bibfnamefont {Y.}~\bibnamefont
  {Kasahara}}, \bibinfo {author} {\bibfnamefont {T.}~\bibnamefont {Kishiume}},
  \bibinfo {author} {\bibfnamefont {T.}~\bibnamefont {Takano}}, \bibinfo
  {author} {\bibfnamefont {K.}~\bibnamefont {Kobayashi}}, \bibinfo {author}
  {\bibfnamefont {E.}~\bibnamefont {Matsuoka}}, \bibinfo {author}
  {\bibfnamefont {H.}~\bibnamefont {Onodera}}, \bibinfo {author} {\bibfnamefont
  {K.}~\bibnamefont {Kuroki}}, \bibinfo {author} {\bibfnamefont
  {Y.}~\bibnamefont {Taguchi}},\ and\ \bibinfo {author} {\bibfnamefont
  {Y.}~\bibnamefont {Iwasa}},\ }\href
  {https://doi.org/10.1103/PhysRevLett.103.077004} {\bibfield  {journal}
  {\bibinfo  {journal} {Phys. Rev. Lett.}\ }\textbf {\bibinfo {volume} {103}},\
  \bibinfo {pages} {077004} (\bibinfo {year} {2009})}\BibitemShut {NoStop}%
\bibitem [{\citenamefont {Pamuk}\ \emph {et~al.}(2017)\citenamefont {Pamuk},
  \citenamefont {Mauri},\ and\ \citenamefont {Calandra}}]{PhysRevB.96.024518}%
  \BibitemOpen
  \bibfield  {author} {\bibinfo {author} {\bibfnamefont {B.}~\bibnamefont
  {Pamuk}}, \bibinfo {author} {\bibfnamefont {F.}~\bibnamefont {Mauri}},\ and\
  \bibinfo {author} {\bibfnamefont {M.}~\bibnamefont {Calandra}},\ }\href
  {https://doi.org/10.1103/PhysRevB.96.024518} {\bibfield  {journal} {\bibinfo
  {journal} {Phys. Rev. B}\ }\textbf {\bibinfo {volume} {96}},\ \bibinfo
  {pages} {024518} (\bibinfo {year} {2017})}\BibitemShut {NoStop}%
\bibitem [{\citenamefont {Tanaka}\ \emph {et~al.}(2022)\citenamefont {Tanaka},
  \citenamefont {Kataoka},\ and\ \citenamefont
  {Yokoya}}]{tanaka_superconductivity_2022}%
  \BibitemOpen
  \bibfield  {author} {\bibinfo {author} {\bibfnamefont {M.}~\bibnamefont
  {Tanaka}}, \bibinfo {author} {\bibfnamefont {N.}~\bibnamefont {Kataoka}},\
  and\ \bibinfo {author} {\bibfnamefont {T.}~\bibnamefont {Yokoya}},\ }\href
  {https://doi.org/10.3390/condmat7020033} {\bibfield  {journal} {\bibinfo
  {journal} {Condensed Matter}\ }\textbf {\bibinfo {volume} {7}},\ \bibinfo
  {pages} {33} (\bibinfo {year} {2022})}\BibitemShut {NoStop}%
\bibitem [{\citenamefont {Crépel}\ and\ \citenamefont
  {Fu}(2022)}]{doi:10.1073/pnas.2117735119}%
  \BibitemOpen
  \bibfield  {author} {\bibinfo {author} {\bibfnamefont {V.}~\bibnamefont
  {Crépel}}\ and\ \bibinfo {author} {\bibfnamefont {L.}~\bibnamefont {Fu}},\
  }\href {https://doi.org/10.1073/pnas.2117735119} {\bibfield  {journal}
  {\bibinfo  {journal} {Proceedings of the National Academy of Sciences}\
  }\textbf {\bibinfo {volume} {119}},\ \bibinfo {pages} {e2117735119} (\bibinfo
  {year} {2022})}\BibitemShut {NoStop}%
\bibitem [{\citenamefont {Crépel}\ and\ \citenamefont
  {Fu}(2021)}]{crepel_new_2021}%
  \BibitemOpen
  \bibfield  {author} {\bibinfo {author} {\bibfnamefont {V.}~\bibnamefont
  {Crépel}}\ and\ \bibinfo {author} {\bibfnamefont {L.}~\bibnamefont {Fu}},\
  }\href {https://doi.org/10.1126/sciadv.abh2233} {\bibfield  {journal}
  {\bibinfo  {journal} {Science Advances}\ }\textbf {\bibinfo {volume} {7}},\
  \bibinfo {pages} {eabh2233} (\bibinfo {year} {2021})}\BibitemShut {NoStop}%
\bibitem [{\citenamefont {Ma}\ \emph {et~al.}(2013)\citenamefont {Ma},
  \citenamefont {Lin},\ and\ \citenamefont {Hu}}]{PhysRevLett.110.107002}%
  \BibitemOpen
  \bibfield  {author} {\bibinfo {author} {\bibfnamefont {T.}~\bibnamefont
  {Ma}}, \bibinfo {author} {\bibfnamefont {H.-Q.}\ \bibnamefont {Lin}},\ and\
  \bibinfo {author} {\bibfnamefont {J.}~\bibnamefont {Hu}},\ }\href
  {https://doi.org/10.1103/PhysRevLett.110.107002} {\bibfield  {journal}
  {\bibinfo  {journal} {Phys. Rev. Lett.}\ }\textbf {\bibinfo {volume} {110}},\
  \bibinfo {pages} {107002} (\bibinfo {year} {2013})}\BibitemShut {NoStop}%
\bibitem [{\citenamefont {Yang}\ \emph {et~al.}(2016)\citenamefont {Yang},
  \citenamefont {Xu}, \citenamefont {Zhang}, \citenamefont {Ma},\ and\
  \citenamefont {Wu}}]{PhysRevB.94.075106}%
  \BibitemOpen
  \bibfield  {author} {\bibinfo {author} {\bibfnamefont {G.}~\bibnamefont
  {Yang}}, \bibinfo {author} {\bibfnamefont {S.}~\bibnamefont {Xu}}, \bibinfo
  {author} {\bibfnamefont {W.}~\bibnamefont {Zhang}}, \bibinfo {author}
  {\bibfnamefont {T.}~\bibnamefont {Ma}},\ and\ \bibinfo {author}
  {\bibfnamefont {C.}~\bibnamefont {Wu}},\ }\href
  {https://doi.org/10.1103/PhysRevB.94.075106} {\bibfield  {journal} {\bibinfo
  {journal} {Phys. Rev. B}\ }\textbf {\bibinfo {volume} {94}},\ \bibinfo
  {pages} {075106} (\bibinfo {year} {2016})}\BibitemShut {NoStop}%
\bibitem [{\citenamefont {Wu}\ \emph {et~al.}(2023)\citenamefont {Wu},
  \citenamefont {Li}, \citenamefont {Wu}, \citenamefont {Hwang},\ and\
  \citenamefont {Cui}}]{wu_electrostatic_2023}%
  \BibitemOpen
  \bibfield  {author} {\bibinfo {author} {\bibfnamefont {Y.}~\bibnamefont
  {Wu}}, \bibinfo {author} {\bibfnamefont {D.}~\bibnamefont {Li}}, \bibinfo
  {author} {\bibfnamefont {C.-L.}\ \bibnamefont {Wu}}, \bibinfo {author}
  {\bibfnamefont {H.~Y.}\ \bibnamefont {Hwang}},\ and\ \bibinfo {author}
  {\bibfnamefont {Y.}~\bibnamefont {Cui}},\ }\href
  {https://doi.org/10.1038/s41578-022-00473-6} {\bibfield  {journal} {\bibinfo
  {journal} {Nature Reviews Materials}\ }\textbf {\bibinfo {volume} {8}},\
  \bibinfo {pages} {41} (\bibinfo {year} {2023})}\BibitemShut {NoStop}%
\bibitem [{\citenamefont {Shen}\ \emph {et~al.}(2020)\citenamefont {Shen},
  \citenamefont {Chu}, \citenamefont {Wu}, \citenamefont {Li}, \citenamefont
  {Wang}, \citenamefont {Zhao}, \citenamefont {Tang}, \citenamefont {Liu},
  \citenamefont {Tian}, \citenamefont {Watanabe}, \citenamefont {Taniguchi},
  \citenamefont {Yang}, \citenamefont {Meng}, \citenamefont {Shi},
  \citenamefont {Yazyev},\ and\ \citenamefont {Zhang}}]{shen_correlated_2020}%
  \BibitemOpen
  \bibfield  {author} {\bibinfo {author} {\bibfnamefont {C.}~\bibnamefont
  {Shen}}, \bibinfo {author} {\bibfnamefont {Y.}~\bibnamefont {Chu}}, \bibinfo
  {author} {\bibfnamefont {Q.}~\bibnamefont {Wu}}, \bibinfo {author}
  {\bibfnamefont {N.}~\bibnamefont {Li}}, \bibinfo {author} {\bibfnamefont
  {S.}~\bibnamefont {Wang}}, \bibinfo {author} {\bibfnamefont {Y.}~\bibnamefont
  {Zhao}}, \bibinfo {author} {\bibfnamefont {J.}~\bibnamefont {Tang}}, \bibinfo
  {author} {\bibfnamefont {J.}~\bibnamefont {Liu}}, \bibinfo {author}
  {\bibfnamefont {J.}~\bibnamefont {Tian}}, \bibinfo {author} {\bibfnamefont
  {K.}~\bibnamefont {Watanabe}}, \bibinfo {author} {\bibfnamefont
  {T.}~\bibnamefont {Taniguchi}}, \bibinfo {author} {\bibfnamefont
  {R.}~\bibnamefont {Yang}}, \bibinfo {author} {\bibfnamefont {Z.~Y.}\
  \bibnamefont {Meng}}, \bibinfo {author} {\bibfnamefont {D.}~\bibnamefont
  {Shi}}, \bibinfo {author} {\bibfnamefont {O.~V.}\ \bibnamefont {Yazyev}},\
  and\ \bibinfo {author} {\bibfnamefont {G.}~\bibnamefont {Zhang}},\ }\href
  {https://doi.org/10.1038/s41567-020-0825-9} {\bibfield  {journal} {\bibinfo
  {journal} {Nature Physics}\ }\textbf {\bibinfo {volume} {16}},\ \bibinfo
  {pages} {520} (\bibinfo {year} {2020})}\BibitemShut {NoStop}%
\bibitem [{\citenamefont {Castro}\ \emph {et~al.}(2007)\citenamefont {Castro},
  \citenamefont {Novoselov}, \citenamefont {Morozov}, \citenamefont {Peres},
  \citenamefont {dos Santos}, \citenamefont {Nilsson}, \citenamefont {Guinea},
  \citenamefont {Geim},\ and\ \citenamefont {Neto}}]{PhysRevLett.99.216802}%
  \BibitemOpen
  \bibfield  {author} {\bibinfo {author} {\bibfnamefont {E.~V.}\ \bibnamefont
  {Castro}}, \bibinfo {author} {\bibfnamefont {K.~S.}\ \bibnamefont
  {Novoselov}}, \bibinfo {author} {\bibfnamefont {S.~V.}\ \bibnamefont
  {Morozov}}, \bibinfo {author} {\bibfnamefont {N.~M.~R.}\ \bibnamefont
  {Peres}}, \bibinfo {author} {\bibfnamefont {J.~M. B.~L.}\ \bibnamefont {dos
  Santos}}, \bibinfo {author} {\bibfnamefont {J.}~\bibnamefont {Nilsson}},
  \bibinfo {author} {\bibfnamefont {F.}~\bibnamefont {Guinea}}, \bibinfo
  {author} {\bibfnamefont {A.~K.}\ \bibnamefont {Geim}},\ and\ \bibinfo
  {author} {\bibfnamefont {A.~H.~C.}\ \bibnamefont {Neto}},\ }\href
  {https://doi.org/10.1103/PhysRevLett.99.216802} {\bibfield  {journal}
  {\bibinfo  {journal} {Phys. Rev. Lett.}\ }\textbf {\bibinfo {volume} {99}},\
  \bibinfo {pages} {216802} (\bibinfo {year} {2007})}\BibitemShut {NoStop}%
\bibitem [{\citenamefont {Zhang}\ \emph {et~al.}(2009)\citenamefont {Zhang},
  \citenamefont {Tang}, \citenamefont {Girit}, \citenamefont {Hao},
  \citenamefont {Martin}, \citenamefont {Zettl}, \citenamefont {Crommie},
  \citenamefont {Shen},\ and\ \citenamefont {Wang}}]{zhang_direct_2009}%
  \BibitemOpen
  \bibfield  {author} {\bibinfo {author} {\bibfnamefont {Y.}~\bibnamefont
  {Zhang}}, \bibinfo {author} {\bibfnamefont {T.-T.}\ \bibnamefont {Tang}},
  \bibinfo {author} {\bibfnamefont {C.}~\bibnamefont {Girit}}, \bibinfo
  {author} {\bibfnamefont {Z.}~\bibnamefont {Hao}}, \bibinfo {author}
  {\bibfnamefont {M.~C.}\ \bibnamefont {Martin}}, \bibinfo {author}
  {\bibfnamefont {A.}~\bibnamefont {Zettl}}, \bibinfo {author} {\bibfnamefont
  {M.~F.}\ \bibnamefont {Crommie}}, \bibinfo {author} {\bibfnamefont {Y.~R.}\
  \bibnamefont {Shen}},\ and\ \bibinfo {author} {\bibfnamefont
  {F.}~\bibnamefont {Wang}},\ }\href {https://doi.org/10.1038/nature08105}
  {\bibfield  {journal} {\bibinfo  {journal} {Nature}\ }\textbf {\bibinfo
  {volume} {459}},\ \bibinfo {pages} {820} (\bibinfo {year}
  {2009})}\BibitemShut {NoStop}%
\bibitem [{\citenamefont {{Hao}}\ \emph {et~al.}(2021)\citenamefont {{Hao}},
  \citenamefont {{Zimmerman}}, \citenamefont {{Ledwith}}, \citenamefont
  {{Khalaf}}, \citenamefont {{Najafabadi}}, \citenamefont {{Watanabe}},
  \citenamefont {{Taniguchi}}, \citenamefont {{Vishwanath}},\ and\
  \citenamefont {{Kim}}}]{2021Sci...371.1133H}%
  \BibitemOpen
  \bibfield  {author} {\bibinfo {author} {\bibfnamefont {Z.}~\bibnamefont
  {{Hao}}}, \bibinfo {author} {\bibfnamefont {A.~M.}\ \bibnamefont
  {{Zimmerman}}}, \bibinfo {author} {\bibfnamefont {P.}~\bibnamefont
  {{Ledwith}}}, \bibinfo {author} {\bibfnamefont {E.}~\bibnamefont {{Khalaf}}},
  \bibinfo {author} {\bibfnamefont {D.~H.}\ \bibnamefont {{Najafabadi}}},
  \bibinfo {author} {\bibfnamefont {K.}~\bibnamefont {{Watanabe}}}, \bibinfo
  {author} {\bibfnamefont {T.}~\bibnamefont {{Taniguchi}}}, \bibinfo {author}
  {\bibfnamefont {A.}~\bibnamefont {{Vishwanath}}},\ and\ \bibinfo {author}
  {\bibfnamefont {P.}~\bibnamefont {{Kim}}},\ }\href
  {https://doi.org/10.1126/science.abg0399} {\bibfield  {journal} {\bibinfo
  {journal} {Science}\ }\textbf {\bibinfo {volume} {371}},\ \bibinfo {pages}
  {1133} (\bibinfo {year} {2021})}\BibitemShut {NoStop}%
\bibitem [{\citenamefont {Lui}\ \emph {et~al.}(2011)\citenamefont {Lui},
  \citenamefont {Li}, \citenamefont {Mak}, \citenamefont {Cappelluti},\ and\
  \citenamefont {Heinz}}]{lui_observation_2011}%
  \BibitemOpen
  \bibfield  {author} {\bibinfo {author} {\bibfnamefont {C.~H.}\ \bibnamefont
  {Lui}}, \bibinfo {author} {\bibfnamefont {Z.}~\bibnamefont {Li}}, \bibinfo
  {author} {\bibfnamefont {K.~F.}\ \bibnamefont {Mak}}, \bibinfo {author}
  {\bibfnamefont {E.}~\bibnamefont {Cappelluti}},\ and\ \bibinfo {author}
  {\bibfnamefont {T.~F.}\ \bibnamefont {Heinz}},\ }\href
  {https://doi.org/10.1038/nphys2102} {\bibfield  {journal} {\bibinfo
  {journal} {Nature Physics}\ }\textbf {\bibinfo {volume} {7}},\ \bibinfo
  {pages} {944} (\bibinfo {year} {2011})}\BibitemShut {NoStop}%
\bibitem [{\citenamefont {Zhou}\ \emph {et~al.}(2021)\citenamefont {Zhou},
  \citenamefont {Xie}, \citenamefont {Ghazaryan}, \citenamefont {Holder},
  \citenamefont {Ehrets}, \citenamefont {Spanton}, \citenamefont {Taniguchi},
  \citenamefont {Watanabe}, \citenamefont {Berg}, \citenamefont {Serbyn},\ and\
  \citenamefont {Young}}]{zhou_half-_2021}%
  \BibitemOpen
  \bibfield  {author} {\bibinfo {author} {\bibfnamefont {H.}~\bibnamefont
  {Zhou}}, \bibinfo {author} {\bibfnamefont {T.}~\bibnamefont {Xie}}, \bibinfo
  {author} {\bibfnamefont {A.}~\bibnamefont {Ghazaryan}}, \bibinfo {author}
  {\bibfnamefont {T.}~\bibnamefont {Holder}}, \bibinfo {author} {\bibfnamefont
  {J.~R.}\ \bibnamefont {Ehrets}}, \bibinfo {author} {\bibfnamefont {E.~M.}\
  \bibnamefont {Spanton}}, \bibinfo {author} {\bibfnamefont {T.}~\bibnamefont
  {Taniguchi}}, \bibinfo {author} {\bibfnamefont {K.}~\bibnamefont {Watanabe}},
  \bibinfo {author} {\bibfnamefont {E.}~\bibnamefont {Berg}}, \bibinfo {author}
  {\bibfnamefont {M.}~\bibnamefont {Serbyn}},\ and\ \bibinfo {author}
  {\bibfnamefont {A.~F.}\ \bibnamefont {Young}},\ }\href
  {https://doi.org/10.1038/s41586-021-03938-w} {\bibfield  {journal} {\bibinfo
  {journal} {Nature}\ }\textbf {\bibinfo {volume} {598}},\ \bibinfo {pages}
  {429} (\bibinfo {year} {2021})}\BibitemShut {NoStop}%
\bibitem [{\citenamefont {Dai}\ \emph {et~al.}(2023)\citenamefont {Dai},
  \citenamefont {Ma}, \citenamefont {Zhang}, \citenamefont {Guo},\ and\
  \citenamefont {Ma}}]{PhysRevB.107.245106}%
  \BibitemOpen
  \bibfield  {author} {\bibinfo {author} {\bibfnamefont {H.}~\bibnamefont
  {Dai}}, \bibinfo {author} {\bibfnamefont {R.}~\bibnamefont {Ma}}, \bibinfo
  {author} {\bibfnamefont {X.}~\bibnamefont {Zhang}}, \bibinfo {author}
  {\bibfnamefont {T.}~\bibnamefont {Guo}},\ and\ \bibinfo {author}
  {\bibfnamefont {T.}~\bibnamefont {Ma}},\ }\href
  {https://doi.org/10.1103/PhysRevB.107.245106} {\bibfield  {journal} {\bibinfo
   {journal} {Phys. Rev. B}\ }\textbf {\bibinfo {volume} {107}},\ \bibinfo
  {pages} {245106} (\bibinfo {year} {2023})}\BibitemShut {NoStop}%
\bibitem [{\citenamefont {Shao}\ \emph {et~al.}(2026)\citenamefont {Shao},
  \citenamefont {Ji}, \citenamefont {Wu}, \citenamefont {Yao},\ and\
  \citenamefont {Yang}}]{shao_possible_2026}%
  \BibitemOpen
  \bibfield  {author} {\bibinfo {author} {\bibfnamefont {Z.-Y.}\ \bibnamefont
  {Shao}}, \bibinfo {author} {\bibfnamefont {J.-H.}\ \bibnamefont {Ji}},
  \bibinfo {author} {\bibfnamefont {C.}~\bibnamefont {Wu}}, \bibinfo {author}
  {\bibfnamefont {D.-X.}\ \bibnamefont {Yao}},\ and\ \bibinfo {author}
  {\bibfnamefont {F.}~\bibnamefont {Yang}},\ }\href
  {https://doi.org/10.1038/s41467-025-67880-5} {\bibfield  {journal} {\bibinfo
  {journal} {Nature Communications}\ }\textbf {\bibinfo {volume} {17}},\
  \bibinfo {pages} {1120} (\bibinfo {year} {2026})}\BibitemShut {NoStop}%
\bibitem [{\citenamefont {Ezawa}(2012)}]{PhysRevLett.109.055502}%
  \BibitemOpen
  \bibfield  {author} {\bibinfo {author} {\bibfnamefont {M.}~\bibnamefont
  {Ezawa}},\ }\href {https://doi.org/10.1103/PhysRevLett.109.055502} {\bibfield
   {journal} {\bibinfo  {journal} {Phys. Rev. Lett.}\ }\textbf {\bibinfo
  {volume} {109}},\ \bibinfo {pages} {055502} (\bibinfo {year}
  {2012})}\BibitemShut {NoStop}%
\bibitem [{\citenamefont {Eek}\ \emph {et~al.}(2025)\citenamefont {Eek},
  \citenamefont {van~'t Westende}, \citenamefont {Klaassen}, \citenamefont
  {Zandvliet}, \citenamefont {Bampoulis},\ and\ \citenamefont
  {Smith}}]{jx2x-fb5b}%
  \BibitemOpen
  \bibfield  {author} {\bibinfo {author} {\bibfnamefont {L.}~\bibnamefont
  {Eek}}, \bibinfo {author} {\bibfnamefont {E.~D.}\ \bibnamefont {van~'t
  Westende}}, \bibinfo {author} {\bibfnamefont {D.~J.}\ \bibnamefont
  {Klaassen}}, \bibinfo {author} {\bibfnamefont {H.~J.~W.}\ \bibnamefont
  {Zandvliet}}, \bibinfo {author} {\bibfnamefont {P.}~\bibnamefont
  {Bampoulis}},\ and\ \bibinfo {author} {\bibfnamefont {C.~M.}\ \bibnamefont
  {Smith}},\ }\href {https://doi.org/10.1103/jx2x-fb5b} {\bibfield  {journal}
  {\bibinfo  {journal} {Phys. Rev. Lett.}\ }\textbf {\bibinfo {volume} {135}},\
  \bibinfo {pages} {206601} (\bibinfo {year} {2025})}\BibitemShut {NoStop}%
\bibitem [{\citenamefont {Nandkishore}\ \emph {et~al.}(2012)\citenamefont
  {Nandkishore}, \citenamefont {Levitov},\ and\ \citenamefont
  {Chubukov}}]{nandkishore_chiral_2012}%
  \BibitemOpen
  \bibfield  {author} {\bibinfo {author} {\bibfnamefont {R.}~\bibnamefont
  {Nandkishore}}, \bibinfo {author} {\bibfnamefont {L.~S.}\ \bibnamefont
  {Levitov}},\ and\ \bibinfo {author} {\bibfnamefont {A.~V.}\ \bibnamefont
  {Chubukov}},\ }\href {https://doi.org/10.1038/nphys2208} {\bibfield
  {journal} {\bibinfo  {journal} {Nature Physics}\ }\textbf {\bibinfo {volume}
  {8}},\ \bibinfo {pages} {158} (\bibinfo {year} {2012})}\BibitemShut {NoStop}%
\bibitem [{\citenamefont {Gonz\'alez}(2008)}]{PhysRevB.78.205431}%
  \BibitemOpen
  \bibfield  {author} {\bibinfo {author} {\bibfnamefont {J.}~\bibnamefont
  {Gonz\'alez}},\ }\href {https://doi.org/10.1103/PhysRevB.78.205431}
  {\bibfield  {journal} {\bibinfo  {journal} {Phys. Rev. B}\ }\textbf {\bibinfo
  {volume} {78}},\ \bibinfo {pages} {205431} (\bibinfo {year}
  {2008})}\BibitemShut {NoStop}%
\bibitem [{\citenamefont {Kiesel}\ \emph {et~al.}(2012)\citenamefont {Kiesel},
  \citenamefont {Platt}, \citenamefont {Hanke}, \citenamefont {Abanin},\ and\
  \citenamefont {Thomale}}]{PhysRevB.86.020507}%
  \BibitemOpen
  \bibfield  {author} {\bibinfo {author} {\bibfnamefont {M.~L.}\ \bibnamefont
  {Kiesel}}, \bibinfo {author} {\bibfnamefont {C.}~\bibnamefont {Platt}},
  \bibinfo {author} {\bibfnamefont {W.}~\bibnamefont {Hanke}}, \bibinfo
  {author} {\bibfnamefont {D.~A.}\ \bibnamefont {Abanin}},\ and\ \bibinfo
  {author} {\bibfnamefont {R.}~\bibnamefont {Thomale}},\ }\href
  {https://doi.org/10.1103/PhysRevB.86.020507} {\bibfield  {journal} {\bibinfo
  {journal} {Phys. Rev. B}\ }\textbf {\bibinfo {volume} {86}},\ \bibinfo
  {pages} {020507} (\bibinfo {year} {2012})}\BibitemShut {NoStop}%
\bibitem [{\citenamefont {Black-Schaffer}\ \emph {et~al.}(2014)\citenamefont
  {Black-Schaffer}, \citenamefont {Wu},\ and\ \citenamefont
  {Le~Hur}}]{PhysRevB.90.054521}%
  \BibitemOpen
  \bibfield  {author} {\bibinfo {author} {\bibfnamefont {A.~M.}\ \bibnamefont
  {Black-Schaffer}}, \bibinfo {author} {\bibfnamefont {W.}~\bibnamefont {Wu}},\
  and\ \bibinfo {author} {\bibfnamefont {K.}~\bibnamefont {Le~Hur}},\ }\href
  {https://doi.org/10.1103/PhysRevB.90.054521} {\bibfield  {journal} {\bibinfo
  {journal} {Phys. Rev. B}\ }\textbf {\bibinfo {volume} {90}},\ \bibinfo
  {pages} {054521} (\bibinfo {year} {2014})}\BibitemShut {NoStop}%
\bibitem [{\citenamefont {Xu}\ \emph {et~al.}(2016)\citenamefont {Xu},
  \citenamefont {Wessel},\ and\ \citenamefont {Meng}}]{PhysRevB.94.115105}%
  \BibitemOpen
  \bibfield  {author} {\bibinfo {author} {\bibfnamefont {X.~Y.}\ \bibnamefont
  {Xu}}, \bibinfo {author} {\bibfnamefont {S.}~\bibnamefont {Wessel}},\ and\
  \bibinfo {author} {\bibfnamefont {Z.~Y.}\ \bibnamefont {Meng}},\ }\href
  {https://doi.org/10.1103/PhysRevB.94.115105} {\bibfield  {journal} {\bibinfo
  {journal} {Phys. Rev. B}\ }\textbf {\bibinfo {volume} {94}},\ \bibinfo
  {pages} {115105} (\bibinfo {year} {2016})}\BibitemShut {NoStop}%
\bibitem [{\citenamefont {Uchoa}\ and\ \citenamefont
  {Castro~Neto}(2007)}]{PhysRevLett.98.146801}%
  \BibitemOpen
  \bibfield  {author} {\bibinfo {author} {\bibfnamefont {B.}~\bibnamefont
  {Uchoa}}\ and\ \bibinfo {author} {\bibfnamefont {A.~H.}\ \bibnamefont
  {Castro~Neto}},\ }\href {https://doi.org/10.1103/PhysRevLett.98.146801}
  {\bibfield  {journal} {\bibinfo  {journal} {Phys. Rev. Lett.}\ }\textbf
  {\bibinfo {volume} {98}},\ \bibinfo {pages} {146801} (\bibinfo {year}
  {2007})}\BibitemShut {NoStop}%
\bibitem [{\citenamefont {Faye}\ \emph {et~al.}(2015)\citenamefont {Faye},
  \citenamefont {Sahebsara},\ and\ \citenamefont
  {S\'en\'echal}}]{PhysRevB.92.085121}%
  \BibitemOpen
  \bibfield  {author} {\bibinfo {author} {\bibfnamefont {J.~P.~L.}\
  \bibnamefont {Faye}}, \bibinfo {author} {\bibfnamefont {P.}~\bibnamefont
  {Sahebsara}},\ and\ \bibinfo {author} {\bibfnamefont {D.}~\bibnamefont
  {S\'en\'echal}},\ }\href {https://doi.org/10.1103/PhysRevB.92.085121}
  {\bibfield  {journal} {\bibinfo  {journal} {Phys. Rev. B}\ }\textbf {\bibinfo
  {volume} {92}},\ \bibinfo {pages} {085121} (\bibinfo {year}
  {2015})}\BibitemShut {NoStop}%
\bibitem [{\citenamefont {Ying}\ and\ \citenamefont
  {Yang}(2020)}]{PhysRevB.102.125125}%
  \BibitemOpen
  \bibfield  {author} {\bibinfo {author} {\bibfnamefont {T.}~\bibnamefont
  {Ying}}\ and\ \bibinfo {author} {\bibfnamefont {S.}~\bibnamefont {Yang}},\
  }\href {https://doi.org/10.1103/PhysRevB.102.125125} {\bibfield  {journal}
  {\bibinfo  {journal} {Phys. Rev. B}\ }\textbf {\bibinfo {volume} {102}},\
  \bibinfo {pages} {125125} (\bibinfo {year} {2020})}\BibitemShut {NoStop}%
\bibitem [{\citenamefont {Honerkamp}(2008)}]{PhysRevLett.100.146404}%
  \BibitemOpen
  \bibfield  {author} {\bibinfo {author} {\bibfnamefont {C.}~\bibnamefont
  {Honerkamp}},\ }\href {https://doi.org/10.1103/PhysRevLett.100.146404}
  {\bibfield  {journal} {\bibinfo  {journal} {Phys. Rev. Lett.}\ }\textbf
  {\bibinfo {volume} {100}},\ \bibinfo {pages} {146404} (\bibinfo {year}
  {2008})}\BibitemShut {NoStop}%
\bibitem [{\citenamefont {Wolf}\ \emph {et~al.}(2022)\citenamefont {Wolf},
  \citenamefont {Gardener}, \citenamefont {Le~Hur},\ and\ \citenamefont
  {Rachel}}]{PhysRevB.105.L100505}%
  \BibitemOpen
  \bibfield  {author} {\bibinfo {author} {\bibfnamefont {S.}~\bibnamefont
  {Wolf}}, \bibinfo {author} {\bibfnamefont {T.}~\bibnamefont {Gardener}},
  \bibinfo {author} {\bibfnamefont {K.}~\bibnamefont {Le~Hur}},\ and\ \bibinfo
  {author} {\bibfnamefont {S.}~\bibnamefont {Rachel}},\ }\href
  {https://doi.org/10.1103/PhysRevB.105.L100505} {\bibfield  {journal}
  {\bibinfo  {journal} {Phys. Rev. B}\ }\textbf {\bibinfo {volume} {105}},\
  \bibinfo {pages} {L100505} (\bibinfo {year} {2022})}\BibitemShut {NoStop}%
\bibitem [{\citenamefont {Blankenbecler}\ \emph {et~al.}(1981)\citenamefont
  {Blankenbecler}, \citenamefont {Scalapino},\ and\ \citenamefont
  {Sugar}}]{PhysRevD.24.2278}%
  \BibitemOpen
  \bibfield  {author} {\bibinfo {author} {\bibfnamefont {R.}~\bibnamefont
  {Blankenbecler}}, \bibinfo {author} {\bibfnamefont {D.~J.}\ \bibnamefont
  {Scalapino}},\ and\ \bibinfo {author} {\bibfnamefont {R.~L.}\ \bibnamefont
  {Sugar}},\ }\href {https://doi.org/10.1103/PhysRevD.24.2278} {\bibfield
  {journal} {\bibinfo  {journal} {Phys. Rev. D}\ }\textbf {\bibinfo {volume}
  {24}},\ \bibinfo {pages} {2278} (\bibinfo {year} {1981})}\BibitemShut
  {NoStop}%
\bibitem [{\citenamefont {Hirsch}(1985)}]{PhysRevB.31.4403}%
  \BibitemOpen
  \bibfield  {author} {\bibinfo {author} {\bibfnamefont {J.~E.}\ \bibnamefont
  {Hirsch}},\ }\href {https://doi.org/10.1103/PhysRevB.31.4403} {\bibfield
  {journal} {\bibinfo  {journal} {Phys. Rev. B}\ }\textbf {\bibinfo {volume}
  {31}},\ \bibinfo {pages} {4403} (\bibinfo {year} {1985})}\BibitemShut
  {NoStop}%
\bibitem [{\citenamefont {Zhang}\ \emph {et~al.}(1997)\citenamefont {Zhang},
  \citenamefont {Carlson},\ and\ \citenamefont
  {Gubernatis}}]{PhysRevB.55.7464}%
  \BibitemOpen
  \bibfield  {author} {\bibinfo {author} {\bibfnamefont {S.~W.}\ \bibnamefont
  {Zhang}}, \bibinfo {author} {\bibfnamefont {J.}~\bibnamefont {Carlson}},\
  and\ \bibinfo {author} {\bibfnamefont {J.~E.}\ \bibnamefont {Gubernatis}},\
  }\href {https://doi.org/10.1103/PhysRevB.55.7464} {\bibfield  {journal}
  {\bibinfo  {journal} {Phys. Rev. B}\ }\textbf {\bibinfo {volume} {55}},\
  \bibinfo {pages} {7464} (\bibinfo {year} {1997})}\BibitemShut {NoStop}%
\bibitem [{\citenamefont {Huang}\ \emph {et~al.}(2001)\citenamefont {Huang},
  \citenamefont {Lin},\ and\ \citenamefont {Gubernatis}}]{PhysRevB.63.115112}%
  \BibitemOpen
  \bibfield  {author} {\bibinfo {author} {\bibfnamefont {Z.~B.}\ \bibnamefont
  {Huang}}, \bibinfo {author} {\bibfnamefont {H.~Q.}\ \bibnamefont {Lin}},\
  and\ \bibinfo {author} {\bibfnamefont {J.~E.}\ \bibnamefont {Gubernatis}},\
  }\href {https://doi.org/10.1103/PhysRevB.63.115112} {\bibfield  {journal}
  {\bibinfo  {journal} {Phys. Rev. B}\ }\textbf {\bibinfo {volume} {63}},\
  \bibinfo {pages} {115112} (\bibinfo {year} {2001})}\BibitemShut {NoStop}%
\bibitem [{\citenamefont {Peres}\ \emph {et~al.}(2004)\citenamefont {Peres},
  \citenamefont {Ara\'ujo},\ and\ \citenamefont {Bozi}}]{PhysRevB.70.195122}%
  \BibitemOpen
  \bibfield  {author} {\bibinfo {author} {\bibfnamefont {N.~M.~R.}\
  \bibnamefont {Peres}}, \bibinfo {author} {\bibfnamefont {M.~A.~N.}\
  \bibnamefont {Ara\'ujo}},\ and\ \bibinfo {author} {\bibfnamefont
  {D.}~\bibnamefont {Bozi}},\ }\href
  {https://doi.org/10.1103/PhysRevB.70.195122} {\bibfield  {journal} {\bibinfo
  {journal} {Phys. Rev. B}\ }\textbf {\bibinfo {volume} {70}},\ \bibinfo
  {pages} {195122} (\bibinfo {year} {2004})}\BibitemShut {NoStop}%
\bibitem [{\citenamefont {Paiva}\ \emph {et~al.}(2005)\citenamefont {Paiva},
  \citenamefont {Scalettar}, \citenamefont {Zheng}, \citenamefont {Singh},\
  and\ \citenamefont {Oitmaa}}]{PhysRevB.72.085123}%
  \BibitemOpen
  \bibfield  {author} {\bibinfo {author} {\bibfnamefont {T.}~\bibnamefont
  {Paiva}}, \bibinfo {author} {\bibfnamefont {R.~T.}\ \bibnamefont
  {Scalettar}}, \bibinfo {author} {\bibfnamefont {W.}~\bibnamefont {Zheng}},
  \bibinfo {author} {\bibfnamefont {R.~R.~P.}\ \bibnamefont {Singh}},\ and\
  \bibinfo {author} {\bibfnamefont {J.}~\bibnamefont {Oitmaa}},\ }\href
  {https://doi.org/10.1103/PhysRevB.72.085123} {\bibfield  {journal} {\bibinfo
  {journal} {Phys. Rev. B}\ }\textbf {\bibinfo {volume} {72}},\ \bibinfo
  {pages} {085123} (\bibinfo {year} {2005})}\BibitemShut {NoStop}%
\bibitem [{\citenamefont {Peres}\ \emph {et~al.}(2006)\citenamefont {Peres},
  \citenamefont {Guinea},\ and\ \citenamefont
  {Castro~Neto}}]{PhysRevB.73.125411}%
  \BibitemOpen
  \bibfield  {author} {\bibinfo {author} {\bibfnamefont {N.~M.~R.}\
  \bibnamefont {Peres}}, \bibinfo {author} {\bibfnamefont {F.}~\bibnamefont
  {Guinea}},\ and\ \bibinfo {author} {\bibfnamefont {A.~H.}\ \bibnamefont
  {Castro~Neto}},\ }\href {https://doi.org/10.1103/PhysRevB.73.125411}
  {\bibfield  {journal} {\bibinfo  {journal} {Phys. Rev. B}\ }\textbf {\bibinfo
  {volume} {73}},\ \bibinfo {pages} {125411} (\bibinfo {year}
  {2006})}\BibitemShut {NoStop}%
\bibitem [{\citenamefont {Castro~Neto}\ \emph {et~al.}(2009)\citenamefont
  {Castro~Neto}, \citenamefont {Guinea}, \citenamefont {Peres}, \citenamefont
  {Novoselov},\ and\ \citenamefont {Geim}}]{RevModPhys.81.109}%
  \BibitemOpen
  \bibfield  {author} {\bibinfo {author} {\bibfnamefont {A.~H.}\ \bibnamefont
  {Castro~Neto}}, \bibinfo {author} {\bibfnamefont {F.}~\bibnamefont {Guinea}},
  \bibinfo {author} {\bibfnamefont {N.~M.~R.}\ \bibnamefont {Peres}}, \bibinfo
  {author} {\bibfnamefont {K.~S.}\ \bibnamefont {Novoselov}},\ and\ \bibinfo
  {author} {\bibfnamefont {A.~K.}\ \bibnamefont {Geim}},\ }\href
  {https://doi.org/10.1103/RevModPhys.81.109} {\bibfield  {journal} {\bibinfo
  {journal} {Rev. Mod. Phys.}\ }\textbf {\bibinfo {volume} {81}},\ \bibinfo
  {pages} {109} (\bibinfo {year} {2009})}\BibitemShut {NoStop}%
\bibitem [{\citenamefont {Qin}\ \emph {et~al.}(2016)\citenamefont {Qin},
  \citenamefont {Shi},\ and\ \citenamefont {Zhang}}]{PhysRevB.94.085103}%
  \BibitemOpen
  \bibfield  {author} {\bibinfo {author} {\bibfnamefont {M.}~\bibnamefont
  {Qin}}, \bibinfo {author} {\bibfnamefont {H.}~\bibnamefont {Shi}},\ and\
  \bibinfo {author} {\bibfnamefont {S.}~\bibnamefont {Zhang}},\ }\href
  {https://doi.org/10.1103/PhysRevB.94.085103} {\bibfield  {journal} {\bibinfo
  {journal} {Phys. Rev. B}\ }\textbf {\bibinfo {volume} {94}},\ \bibinfo
  {pages} {085103} (\bibinfo {year} {2016})}\BibitemShut {NoStop}%
\bibitem [{\citenamefont {Xu}\ \emph {et~al.}(2022)\citenamefont {Xu},
  \citenamefont {Shi}, \citenamefont {Vitali}, \citenamefont {Qin},\ and\
  \citenamefont {Zhang}}]{PhysRevResearch.4.013239}%
  \BibitemOpen
  \bibfield  {author} {\bibinfo {author} {\bibfnamefont {H.}~\bibnamefont
  {Xu}}, \bibinfo {author} {\bibfnamefont {H.}~\bibnamefont {Shi}}, \bibinfo
  {author} {\bibfnamefont {E.}~\bibnamefont {Vitali}}, \bibinfo {author}
  {\bibfnamefont {M.}~\bibnamefont {Qin}},\ and\ \bibinfo {author}
  {\bibfnamefont {S.}~\bibnamefont {Zhang}},\ }\href
  {https://doi.org/10.1103/PhysRevResearch.4.013239} {\bibfield  {journal}
  {\bibinfo  {journal} {Phys. Rev. Res.}\ }\textbf {\bibinfo {volume} {4}},\
  \bibinfo {pages} {013239} (\bibinfo {year} {2022})}\BibitemShut {NoStop}%
\bibitem [{\citenamefont {Hirsch}\ and\ \citenamefont
  {Lin}(1988)}]{PhysRevB.37.5070}%
  \BibitemOpen
  \bibfield  {author} {\bibinfo {author} {\bibfnamefont {J.~E.}\ \bibnamefont
  {Hirsch}}\ and\ \bibinfo {author} {\bibfnamefont {H.~Q.}\ \bibnamefont
  {Lin}},\ }\href {https://doi.org/10.1103/PhysRevB.37.5070} {\bibfield
  {journal} {\bibinfo  {journal} {Phys. Rev. B}\ }\textbf {\bibinfo {volume}
  {37}},\ \bibinfo {pages} {5070} (\bibinfo {year} {1988})}\BibitemShut
  {NoStop}%
\bibitem [{\citenamefont {Lin}\ \emph {et~al.}(1988)\citenamefont {Lin},
  \citenamefont {Hirsch},\ and\ \citenamefont {Scalapino}}]{PhysRevB.37.7359}%
  \BibitemOpen
  \bibfield  {author} {\bibinfo {author} {\bibfnamefont {H.~Q.}\ \bibnamefont
  {Lin}}, \bibinfo {author} {\bibfnamefont {J.~E.}\ \bibnamefont {Hirsch}},\
  and\ \bibinfo {author} {\bibfnamefont {D.~J.}\ \bibnamefont {Scalapino}},\
  }\href {https://doi.org/10.1103/PhysRevB.37.7359} {\bibfield  {journal}
  {\bibinfo  {journal} {Phys. Rev. B}\ }\textbf {\bibinfo {volume} {37}},\
  \bibinfo {pages} {7359} (\bibinfo {year} {1988})}\BibitemShut {NoStop}%
\bibitem [{\citenamefont {Ma}\ \emph {et~al.}(2011)\citenamefont {Ma},
  \citenamefont {Huang}, \citenamefont {Hu},\ and\ \citenamefont
  {Lin}}]{PhysRevB.84.121410}%
  \BibitemOpen
  \bibfield  {author} {\bibinfo {author} {\bibfnamefont {T.}~\bibnamefont
  {Ma}}, \bibinfo {author} {\bibfnamefont {Z.}~\bibnamefont {Huang}}, \bibinfo
  {author} {\bibfnamefont {F.}~\bibnamefont {Hu}},\ and\ \bibinfo {author}
  {\bibfnamefont {H.-Q.}\ \bibnamefont {Lin}},\ }\href
  {https://doi.org/10.1103/PhysRevB.84.121410} {\bibfield  {journal} {\bibinfo
  {journal} {Phys. Rev. B}\ }\textbf {\bibinfo {volume} {84}},\ \bibinfo
  {pages} {121410} (\bibinfo {year} {2011})}\BibitemShut {NoStop}%
\bibitem [{\citenamefont {Jiang}\ \emph {et~al.}(2008)\citenamefont {Jiang},
  \citenamefont {Yao}, \citenamefont {Carlson}, \citenamefont {Chen},\ and\
  \citenamefont {Hu}}]{PhysRevB.77.235420}%
  \BibitemOpen
  \bibfield  {author} {\bibinfo {author} {\bibfnamefont {Y.}~\bibnamefont
  {Jiang}}, \bibinfo {author} {\bibfnamefont {D.-X.}\ \bibnamefont {Yao}},
  \bibinfo {author} {\bibfnamefont {E.~W.}\ \bibnamefont {Carlson}}, \bibinfo
  {author} {\bibfnamefont {H.-D.}\ \bibnamefont {Chen}},\ and\ \bibinfo
  {author} {\bibfnamefont {J.}~\bibnamefont {Hu}},\ }\href
  {https://doi.org/10.1103/PhysRevB.77.235420} {\bibfield  {journal} {\bibinfo
  {journal} {Phys. Rev. B}\ }\textbf {\bibinfo {volume} {77}},\ \bibinfo
  {pages} {235420} (\bibinfo {year} {2008})}\BibitemShut {NoStop}%
\bibitem [{\citenamefont {Scalapino}(1995)}]{Scalapino1995TheCF}%
  \BibitemOpen
  \bibfield  {author} {\bibinfo {author} {\bibfnamefont {D.~J.}\ \bibnamefont
  {Scalapino}},\ }\href {https://api.semanticscholar.org/CorpusID:43967680}
  {\bibfield  {journal} {\bibinfo  {journal} {Physics Reports}\ }\textbf
  {\bibinfo {volume} {250}},\ \bibinfo {pages} {329} (\bibinfo {year}
  {1995})}\BibitemShut {NoStop}%
\bibitem [{\citenamefont {Scalettar}\ \emph {et~al.}(1989)\citenamefont
  {Scalettar}, \citenamefont {Loh}, \citenamefont {Gubernatis}, \citenamefont
  {Moreo}, \citenamefont {White}, \citenamefont {Scalapino}, \citenamefont
  {Sugar},\ and\ \citenamefont {Dagotto}}]{PhysRevLett.62.1407}%
  \BibitemOpen
  \bibfield  {author} {\bibinfo {author} {\bibfnamefont {R.~T.}\ \bibnamefont
  {Scalettar}}, \bibinfo {author} {\bibfnamefont {E.~Y.}\ \bibnamefont {Loh}},
  \bibinfo {author} {\bibfnamefont {J.~E.}\ \bibnamefont {Gubernatis}},
  \bibinfo {author} {\bibfnamefont {A.}~\bibnamefont {Moreo}}, \bibinfo
  {author} {\bibfnamefont {S.~R.}\ \bibnamefont {White}}, \bibinfo {author}
  {\bibfnamefont {D.~J.}\ \bibnamefont {Scalapino}}, \bibinfo {author}
  {\bibfnamefont {R.~L.}\ \bibnamefont {Sugar}},\ and\ \bibinfo {author}
  {\bibfnamefont {E.}~\bibnamefont {Dagotto}},\ }\href
  {https://doi.org/10.1103/PhysRevLett.62.1407} {\bibfield  {journal} {\bibinfo
   {journal} {Phys. Rev. Lett.}\ }\textbf {\bibinfo {volume} {62}},\ \bibinfo
  {pages} {1407} (\bibinfo {year} {1989})}\BibitemShut {NoStop}%
\bibitem [{\citenamefont {Scalapino}(2012)}]{RevModPhys.84.1383}%
  \BibitemOpen
  \bibfield  {author} {\bibinfo {author} {\bibfnamefont {D.~J.}\ \bibnamefont
  {Scalapino}},\ }\href {https://doi.org/10.1103/RevModPhys.84.1383} {\bibfield
   {journal} {\bibinfo  {journal} {Rev. Mod. Phys.}\ }\textbf {\bibinfo
  {volume} {84}},\ \bibinfo {pages} {1383} (\bibinfo {year}
  {2012})}\BibitemShut {NoStop}%
\bibitem [{\citenamefont {Loh}\ \emph {et~al.}(1990)\citenamefont {Loh},
  \citenamefont {Gubernatis}, \citenamefont {Scalettar}, \citenamefont {White},
  \citenamefont {Scalapino},\ and\ \citenamefont {Sugar}}]{PhysRevB.41.9301}%
  \BibitemOpen
  \bibfield  {author} {\bibinfo {author} {\bibfnamefont {E.~Y.}\ \bibnamefont
  {Loh}}, \bibinfo {author} {\bibfnamefont {J.~E.}\ \bibnamefont {Gubernatis}},
  \bibinfo {author} {\bibfnamefont {R.~T.}\ \bibnamefont {Scalettar}}, \bibinfo
  {author} {\bibfnamefont {S.~R.}\ \bibnamefont {White}}, \bibinfo {author}
  {\bibfnamefont {D.~J.}\ \bibnamefont {Scalapino}},\ and\ \bibinfo {author}
  {\bibfnamefont {R.~L.}\ \bibnamefont {Sugar}},\ }\href
  {https://doi.org/10.1103/PhysRevB.41.9301} {\bibfield  {journal} {\bibinfo
  {journal} {Phys. Rev. B}\ }\textbf {\bibinfo {volume} {41}},\ \bibinfo
  {pages} {9301} (\bibinfo {year} {1990})}\BibitemShut {NoStop}%
\bibitem [{\citenamefont {Zhang}\ \emph {et~al.}(1995)\citenamefont {Zhang},
  \citenamefont {Carlson},\ and\ \citenamefont
  {Gubernatis}}]{PhysRevLett.74.3652}%
  \BibitemOpen
  \bibfield  {author} {\bibinfo {author} {\bibfnamefont {S.}~\bibnamefont
  {Zhang}}, \bibinfo {author} {\bibfnamefont {J.}~\bibnamefont {Carlson}},\
  and\ \bibinfo {author} {\bibfnamefont {J.~E.}\ \bibnamefont {Gubernatis}},\
  }\href {https://doi.org/10.1103/PhysRevLett.74.3652} {\bibfield  {journal}
  {\bibinfo  {journal} {Phys. Rev. Lett.}\ }\textbf {\bibinfo {volume} {74}},\
  \bibinfo {pages} {3652} (\bibinfo {year} {1995})}\BibitemShut {NoStop}%
\bibitem [{\citenamefont {Cai}\ \emph {et~al.}(2026)\citenamefont {Cai},
  \citenamefont {Xiong}, \citenamefont {Liang},\ and\ \citenamefont
  {Ma}}]{cai2026data}%
  \BibitemOpen
  \bibfield  {author} {\bibinfo {author} {\bibfnamefont {Y.}~\bibnamefont
  {Cai}}, \bibinfo {author} {\bibfnamefont {Y.}~\bibnamefont {Xiong}}, \bibinfo
  {author} {\bibfnamefont {Y.}~\bibnamefont {Liang}},\ and\ \bibinfo {author}
  {\bibfnamefont {T.}~\bibnamefont {Ma}},\ }\href@noop {} {\bibinfo {title}
  {Data associated with {Tuning} superconducting pairing symmetry via a
  staggered potential in the doped honeycomb {Hubbard} model {[Dataset]}}},\
  \bibinfo {howpublished} {Zenodo} (\bibinfo {year} {2026}),\ \bibinfo {note}
  {{\hypersetup{urlcolor=magenta}\href{https://doi.org/10.5281/zenodo.21261572}{https://doi.org/10.5281/zenodo.21261572}}}\BibitemShut
  {NoStop}%
\bibitem [{\citenamefont {Xiong}\ \emph {et~al.}(2026)\citenamefont {Xiong},
  \citenamefont {Cai},\ and\ \citenamefont {Ma}}]{1sgg-ztw8}%
  \BibitemOpen
  \bibfield  {author} {\bibinfo {author} {\bibfnamefont {Y.}~\bibnamefont
  {Xiong}}, \bibinfo {author} {\bibfnamefont {Y.}~\bibnamefont {Cai}},\ and\
  \bibinfo {author} {\bibfnamefont {T.}~\bibnamefont {Ma}},\ }\href
  {https://doi.org/10.1103/1sgg-ztw8} {\bibfield  {journal} {\bibinfo
  {journal} {Phys. Rev. B}\ }\textbf {\bibinfo {volume} {113}},\ \bibinfo
  {pages} {125134} (\bibinfo {year} {2026})}\BibitemShut {NoStop}%
\bibitem [{\citenamefont {Xu}\ \emph {et~al.}(2024)\citenamefont {Xu},
  \citenamefont {Chung}, \citenamefont {Qin}, \citenamefont {Schollwöck},
  \citenamefont {White},\ and\ \citenamefont {Zhang}}]{xu_coexistence_2024}%
  \BibitemOpen
  \bibfield  {author} {\bibinfo {author} {\bibfnamefont {H.}~\bibnamefont
  {Xu}}, \bibinfo {author} {\bibfnamefont {C.-M.}\ \bibnamefont {Chung}},
  \bibinfo {author} {\bibfnamefont {M.}~\bibnamefont {Qin}}, \bibinfo {author}
  {\bibfnamefont {U.}~\bibnamefont {Schollwöck}}, \bibinfo {author}
  {\bibfnamefont {S.~R.}\ \bibnamefont {White}},\ and\ \bibinfo {author}
  {\bibfnamefont {S.}~\bibnamefont {Zhang}},\ }\href
  {https://doi.org/10.1126/science.adh7691} {\bibfield  {journal} {\bibinfo
  {journal} {Science}\ }\textbf {\bibinfo {volume} {384}},\ \bibinfo {pages}
  {eadh7691} (\bibinfo {year} {2024})}\BibitemShut {NoStop}%
\bibitem [{\citenamefont {Chen}\ \emph {et~al.}(2020)\citenamefont {Chen},
  \citenamefont {Chu}, \citenamefont {Huang},\ and\ \citenamefont
  {Ma}}]{PhysRevB.101.155413}%
  \BibitemOpen
  \bibfield  {author} {\bibinfo {author} {\bibfnamefont {W.}~\bibnamefont
  {Chen}}, \bibinfo {author} {\bibfnamefont {Y.}~\bibnamefont {Chu}}, \bibinfo
  {author} {\bibfnamefont {T.}~\bibnamefont {Huang}},\ and\ \bibinfo {author}
  {\bibfnamefont {T.}~\bibnamefont {Ma}},\ }\href
  {https://doi.org/10.1103/PhysRevB.101.155413} {\bibfield  {journal} {\bibinfo
   {journal} {Phys. Rev. B}\ }\textbf {\bibinfo {volume} {101}},\ \bibinfo
  {pages} {155413} (\bibinfo {year} {2020})}\BibitemShut {NoStop}%
\end{thebibliography}%

\end{document}